\documentclass[%
 aip,
 amsmath,amssymb,
 reprint,%
]{revtex4-1}

\usepackage{graphicx}% Include figure files
\usepackage{dcolumn}% Align table columns on decimal point
\usepackage{bm}% bold math
\usepackage[utf8]{inputenc}
\usepackage[T1]{fontenc}
\usepackage{mathptmx}
\usepackage{etoolbox}

\long\def\comment#1{}
\makeatletter
\def\@email#1#2{%
 \endgroup
 \patchcmd{\titleblock@produce}
  {\frontmatter@RRAPformat}
  {\frontmatter@RRAPformat{\produce@RRAP{*#1\href{mailto:#2}{#2}}}\frontmatter@RRAPformat}
  {}{}
}%
\makeatother
\begin{document}

\preprint{AIP/123-QED}

\title{Relaminarization of turbulent puffs in pipe flow}
% Force line breaks with \\
\author{Basheer A. Khan}
 \altaffiliation[Graduated from the Indian Institute of Technology (I.I.T.) Kanpur.
 Currently a postdoctoral fellow at the ]{Department of Mechanical Engineering,
 Ben-Gurion University, Israel.}%Lines break automatically or can be forced with \\
\author{Shai Arogeti}%
 %\email{Second.Author@institution.edu.}
%
\author{Oriel Shoshani}
\author{Alexander Yakhot}
\altaffiliation{
 Corresponding author, yakhot@bgu.ac.il %\\This line break forced with \textbackslash\textbackslash
}%
\affiliation{Department of Mechanical Engineering,
 Ben-Gurion University, Israel.
}%\\This line break forced with \textbackslash\textbackslash

\date{\today}% It is always \today, today,

\begin{abstract}
This study examines the relaminarization of turbulent puffs in pipe flow using highly resolved direct numerical simulations
at Reynolds numbers of 1880, 1900, and 1920. The exponential decay of the total energy of streamwise velocity fluctuations
and the weak dependence of the decay rate on the Reynolds number were verified. In the cross-section of the laminar-turbulent
interface at the trailing edge of the puff, an analysis of the spatio-temporal evolution of the streamline patterns reveals
a complex topology with saddle-node pairs. In this case, the saddle and the node move toward each other during relaminarization
until they collide and vanish. By tracing the vanishing saddle-node pairs over time, we discovered that the distance between
the saddle and the node scales as a square root function of time, a generic characteristic of a time-evolving saddle-node bifurcation.
\end{abstract}

\maketitle
%%%%%%%%%%%%%%%%%%%%
%\input IntroR1-V2.tex
%\input comp_setupR1.tex
%\input exp_decayR1-V2.tex
%\input ISNP.tex
%\input ConclusionsR1-V3.tex
%%%%%%%%%%%%%%%%%%
\section{INTRODUCTION}\label{sec: intro}
Turbulence in a pipe appears as "puffs",  which are localized, self-sustaining, and confined patches\cite{Wygi1973} that occur at
Reynolds numbers ranging between 1700 and 2000. This makes them intriguing subjects for studying the rapid transition from
laminar to turbulent flow at the upstream edge, and the prolonged recovery from turbulent to laminar flow downstream
(see references \onlinecite{Barkley2016, Avila2023}, and references therein). Thirty years after turbulent puffs were
discovered, it was found that they are not sustained for an indefinite time. Breakthrough measurements\cite{Hof2006,Peixinho2006}
followed by numerical simulations\cite{Willis2007, Moxey2011} revealed unexpected results that, if the Reynolds number increases,
puffs survive for a long time before decaying.
Fifty years after the discovery of turbulent puffs in pipes, it has now been established that they represent {\it metastable} states,
in that they occur as a stable, self-sustaining, equilibrium structure with a finite lifetime, but eventually and abruptly move
to a more favored (laminar) state\cite{Avila2023}. The term ``equilibrium lifetime'' (hereafter referred to as ``lifetime of
equilibrium state'') is a statistical quantity that must be estimated in terms of probability.
Notably, the probability that a turbulent puff remains in equilibrium after time $t$ exhibits an exponential decay
$P(t)\sim\exp(-Ct)$, with $C=C(Re)$ referred to as ``the rate of escape from the turbulent state''\cite{Faisst2004}
or ``the rate of survival probability'' using the term ``survival probability''\cite{AltmannPortelaTel2013, Tel2015}.
In Ref. \onlinecite{Faisst2004}, the relaminarization of turbulent puff was linked to a particle in a box with a hole,
where the chaotic dynamics inside a box ends upon a particle leaves through a hole. This analogy is similar to that of a billiard
table with a hole and a chaotically moving ball, which mathematicians have extensively studied as an example of an open dynamic
system\cite{PianigianiYorke1979, AttarchiBunimovich2020}. A ``hole'' (or leak) occurs in a wide variety of physical scenarios,
from acoustics to plasma physics. There are examples of open (leaking) chaotic systems, whose treatment is based on the
transient (finite-time) chaos theory. Considering different examples, expressions of survival probability have been derived as
an exponential function of time\cite{AltmannPortelaTel2013, Tel2015}.

Puff splitting was an additional concern requiring thorough investigation as the Reynolds number increases. Studies of
the critical Reynolds number at which the splitting occurs were resolved by the findings reported in Ref. \onlinecite{Avila2011}.
Measurements and direct numerical simulation (DNS) results were approximated by a single superexponential fit,
$\tau=\exp[\exp(aRe+b)]$, for the mean equilibrium lifetime of a puff before sudden decaying or splitting. This indicates
approximately 2040 as a critical $Re$-limit for the most probable occurrence of a localized puff in a pipe. Therefore, below
the critical value $Re=2040$, the decay exhibits a mean lifetime that depends on $Re$ and is longer than the mean lifetime before
the decaying event\cite{Avila2011}. In particular, the
equilibrium lifetime of turbulent puffs varies from a few hundred to approximately $2\times 10^7$ advective time
units ($D/U_m$) in
the Reynolds number range $Re$=1720--2040. In any case, the puffs do not persist indefinitely; their mean lifetime at
equilibrium is finite for any given $Re$, after which the turbulence decays and the flow relaminarizes. The upper limit of
$Re_c$=2040 corresponds to experimentally inaccessible pipes with a length of approximately $2\times 10^7$ diameters.
The equilibrium lifetimes up to $Re$=2000 have been determined using high-performance computing resources for more
than 1000 independent DNS runs of pipe flow\cite{NemotoAlexakis2021}.

In the fifty years since puffs were discovered, a great deal of information has been gathered, with an emphasis on the
self-sustaining mechanism, estimates of the equilibrium lifetime, and the underlying cause for their formation.
Understanding the
self-maintenance of puffs made it possible to destroy it with subsequent relaminarization\cite{Hof2010, Kuhnen2018a}.
Additionally, it has been established that turbulent puffs eventually relaminarize independently, on their own, following
an abrupt departure
from an equilibrium chaotic state. This may occur as a result of the self-sustaining mechanism breakdown, but its
cause is still unclear. After the sudden onset, the decay of the energy of the longitudinal and transverse turbulent velocity
components occurs differently. The transverse motion components are responsible for the downstream vorticity ($\omega_z$) of
small eddies, which in turn transfer energy from the mean flow shear to the turbulent eddies. The transverse motion of small
eddies  decays much faster\cite{KhanArogetiYakhot2024}, and, therefore, large-scale longitudinal eddies remain without any
amplification mechanism. The scenario is analogous to the DNS results of the plane Couette flow\cite{Chantry2014}, where
the viscous decay of streaks causes the exponential decay. The key distinction is that puffs have a persistent size, only
weakly dependent on Reynolds number size, which serves as a characteristic length. This, in turn, implies the presence of
a characteristic decay rate. Accordingly, mean lifetimes are the most suitable quantities to characterize puffs at a given
Re. The mean decay rate, which describes their relaminarization, is the subject of this study.

The crisis Reynolds number was conventionally introduced as the minimum above which turbulence persists for $t > 1000$
advective time units
$(D/U_m)$; it was experimentally\cite{Peixinho2006} and numerically\cite{Willis2007} estimated as $Re=1750$ and $1870$,
respectively.  At Reynolds numbers $Re \sim 1900$, the mean lifetime of an equilibrium (localized) turbulent puff before decay
is $O(1000)$ and in the range Re=1780--1840 it is $O(100)$ time units $(D/U_m)$\cite{Avila2011}. In Ref.
\onlinecite{KhanArogetiYakhot2024}, we conducted DNS for a range of sub-crisis Reynolds numbers Re=1720--1840. The DNS results suggest that the relaminarization of a puff is governed by the exponential decay of energy of the streamwise velocity component fluctuations, $e_z \propto 10^{-\kappa\tau},~\tau=tU_m/D$.
In 1979, Sreenivasan\cite{NarasimhaSreeni1979} used a cubic expression for fitting the rate of decay $\kappa=B(d-Re)^3$
($B$ and $d$ are constants). This expression turned out to be inapplicable for the relaminarization of turbulent puffs.
We have shown that using a cubic expression poses an ill-conditioned problem, and the inclusion of a constant is required
to resolve this issue, namely, $\kappa=B(d-Re)^3+C$ offers a good fit to the DNS results for the low Reynolds number range of Re=1720-1840
\cite{KhanArogetiYakhot2024}. In the current study, we check this approximation for $Re$=1880, 1900 and 1920.

Although substantial progress has been achieved in understanding the onset and persistence of turbulent puffs in pipe flow,
the mechanisms that underlie relaminarization remain unresolved.
Our objective is to monitor turbulent puffs from their self-sustaining equilibrium state to complete relaminarization.
In particular, we investigate the spatiotemporal evolution of organized flow
structures, with an emphasis on the sectional streamlines and near-wall streaks in the transition region, near the trailing
edge of the puff.

The remainder of this paper is organized as follows. Section \ref{sec: setup} details the computational setup and numerical 
methodology used for DNS of transitional pipe flow. Section \ref{sec: res} presents the main results. 
Subsection \ref{subsec: expdecay} discusses the decay of a turbulent puff after an abrupt escape from an equilibrium turbulent 
state. Subsections \ref{subsec: IPoints} and \ref{subsec: SNpoints} analyze the spatio-temporal evolution of instantaneous 
streamwise velocity profiles and the topology of sectional streamline patterns. In particular, the appearance of saddle and 
nodal points near the laminar–turbulent interface, which exhibits intricate topological features such as saddle-node bifurcation 
at the very end of relaminarization.

\section{COMPUTATIONAL SET-UP}\label{sec: setup}
Direct numerical simulations (DNS) of laminar-turbulent transitional flow in a pipe were performed using the Openpipeflow 
Navier-Stokes solver\cite[]{Willis2017}. The solver is a spectral finite-difference CFD code with centerline velocity of laminar 
flow ($U_c$), pipe radius ($R$), and time ($R/U_c$) as dimensional scales. The bulk (mean) flow was fixed; therefore, 
$U_c=2U_m$, and consequently, $Re=U_cR/\nu=U_mD/\nu$, where $U_m$ is mean velocity. The length of the pipe was set to $16\pi D$
(hereafter referred to as $L=50D$) to ensure that a puff is surrounded by laminar flow. The flow is periodic in the axial 
direction, and  velocity field was expanded in a Fourier series in the azimuthal ($\theta$) and streamwise ($z$) directions. 
Finite-difference discretization with $r_n$ points as Chebyshev polynomial roots were used in the radial direction. Accordingly, 
the $\theta \times r \times z$ resolution consists of $72 \times 64 \times 1152$ grid points. The simulations have been carried 
with a time step $\Delta t=$0.0025 in $D/U_m$ units. Every 50 times steps, the data were collected 
inside an $8D$-width window centered around the location $z=0$, where the instantaneous transverse motion energy 
$e_{\bot}=\int\int(u_r^2+u_{\theta}^2)r\text{d}r\text{d}\theta$ was found to be the highest; hereafter, $z$ denotes the location 
in the reference frame linked to the moving puff. The simulations were started from snapshots of a localized turbulent puff and 
continued until complete relaminarization, that is, the centerline volume-average velocity is greater than 0.999, which is a 
heuristic threshold in the Openpipeflow code version that we used. 
\section{Results}\label{sec: res}
In the present study, we performed highly resolved DNS runs for Reynolds numbers of 1880, 1900, and 1920. The 22 initial conditions for each $Re$ were chosen from statistical equilibrium states that were achieved at a different Reynolds number. In the range $1750<Re<2040$, the mean equilibrium lifetimes of a puff before it decays or splits were approximated by
a superexponential fit\cite{Moxey2011}:  $t_l=\exp[\exp(0.005556Re-8.499)]$; here, and hereafter, the time is in
advective ($D/U_m$) time units. In the current study, the mean lifetimes until complete relaminarization are $t_l$=780,~2100, and 5950 for $Re=$1880,~1900, and 1920, respectively. At the same time, the spread of $t_l$ for $Re$=1880,~1900, and 1920 was significant, with lifetime ranges of $186<t_l<3010,~225<t_l<8770,\text{ and }600<t_l<19050$,
respectively. This is consistent with the findings of Ref. \onlinecite{Faisst2004}, which argue that the abrupt exit from a turbulent (chaotic) state followed by relaminarization is a global phenomenon, with the lifetime depending significantly on the initial condition.
\subsection{Exponential decay}\label{subsec: expdecay}
\begin{figure*}
{\centering(a)}
\centerline{
    \includegraphics[width=0.65 \textwidth]{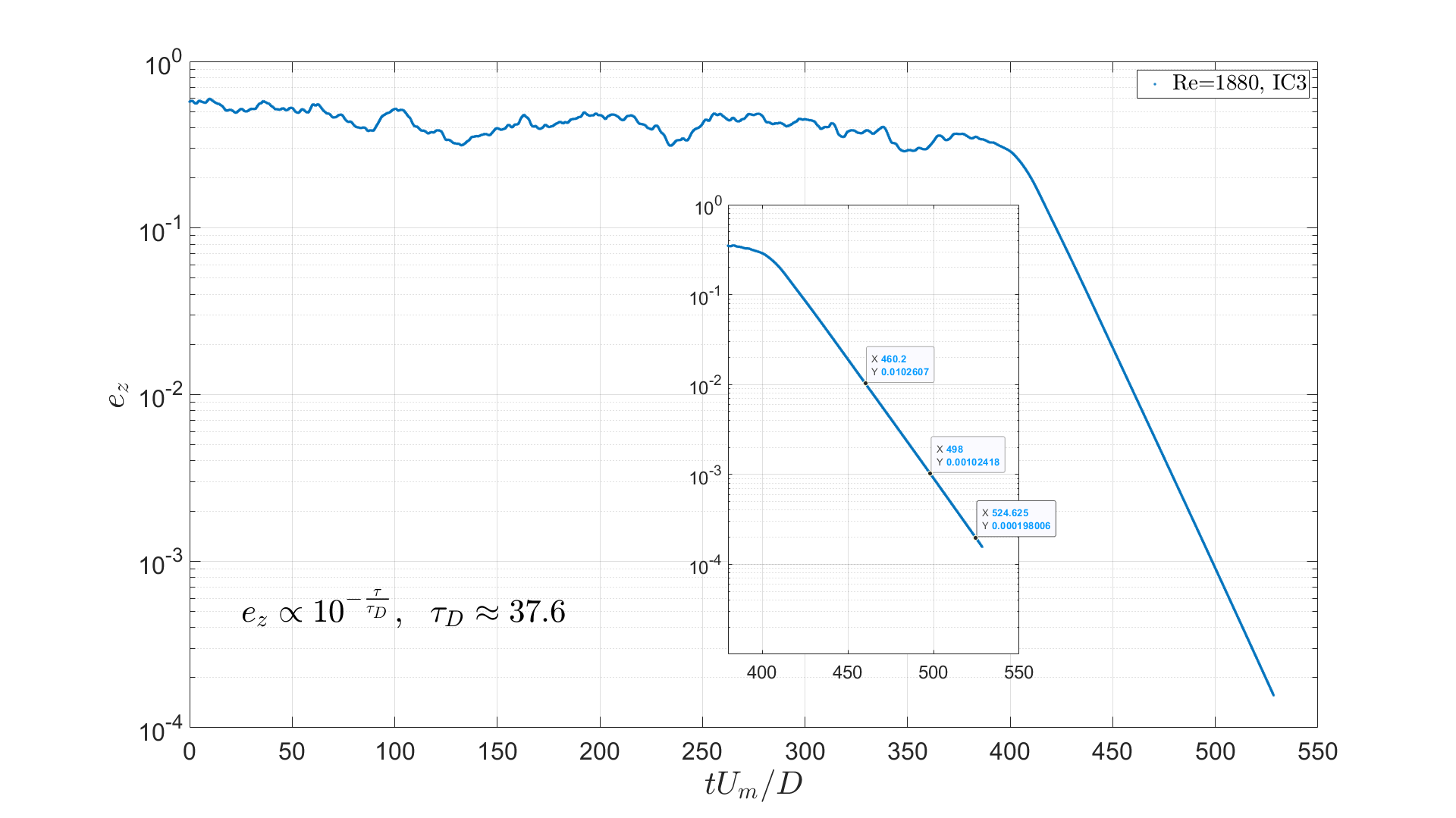}
}
{\centering(b)}
\centerline{
    %{\includegraphics[width=0.65 \textwidth]{Re1900_IC1_ez_tauDc}}
    {\includegraphics[width=0.65 \textwidth]{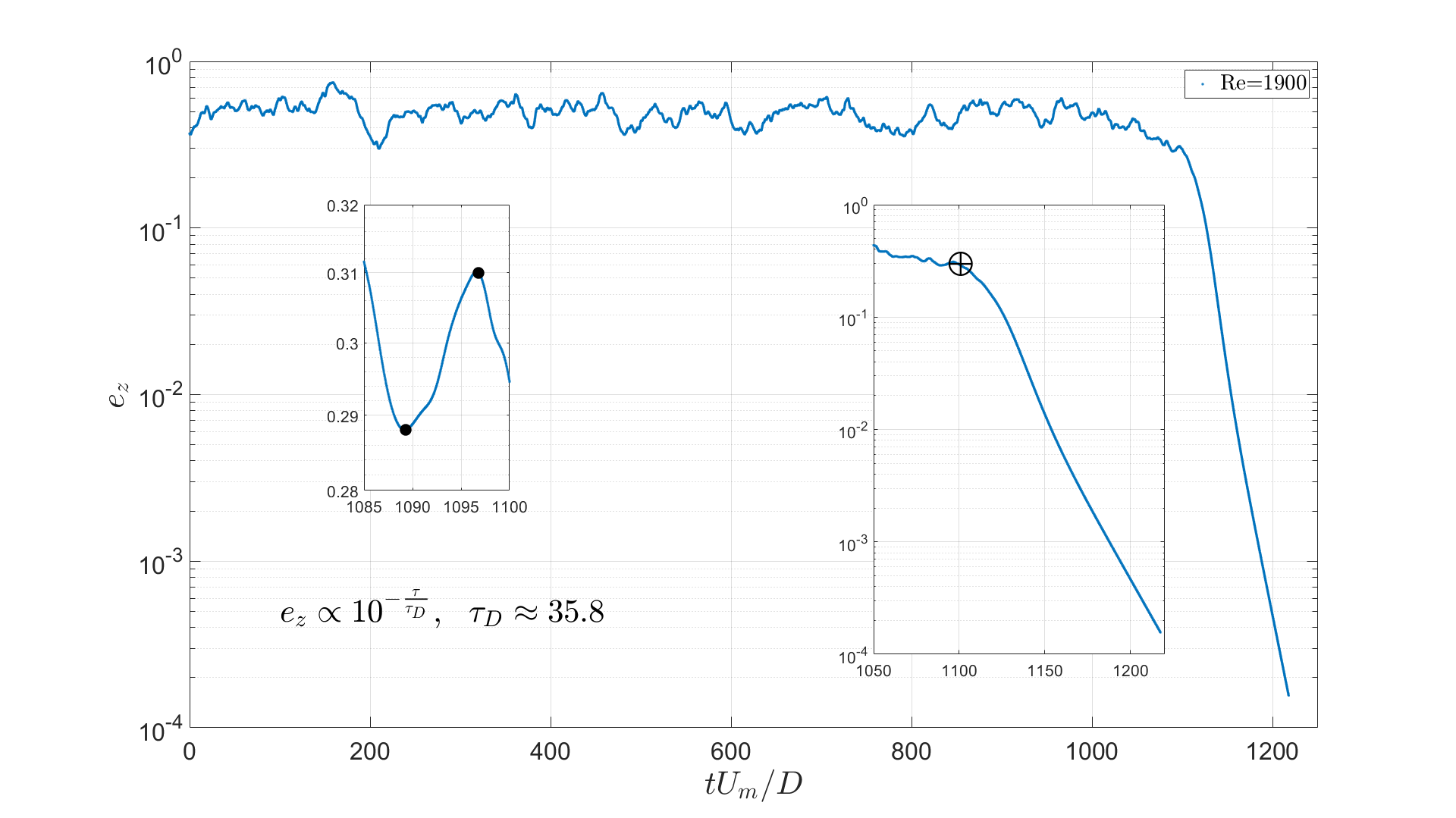}}
}
{\centering(c)}
\centerline{
    %{\includegraphics[width=0.85 \textwidth]{Re1920_IC3_ez_tauD}}
    {\includegraphics[width=0.65 \textwidth]{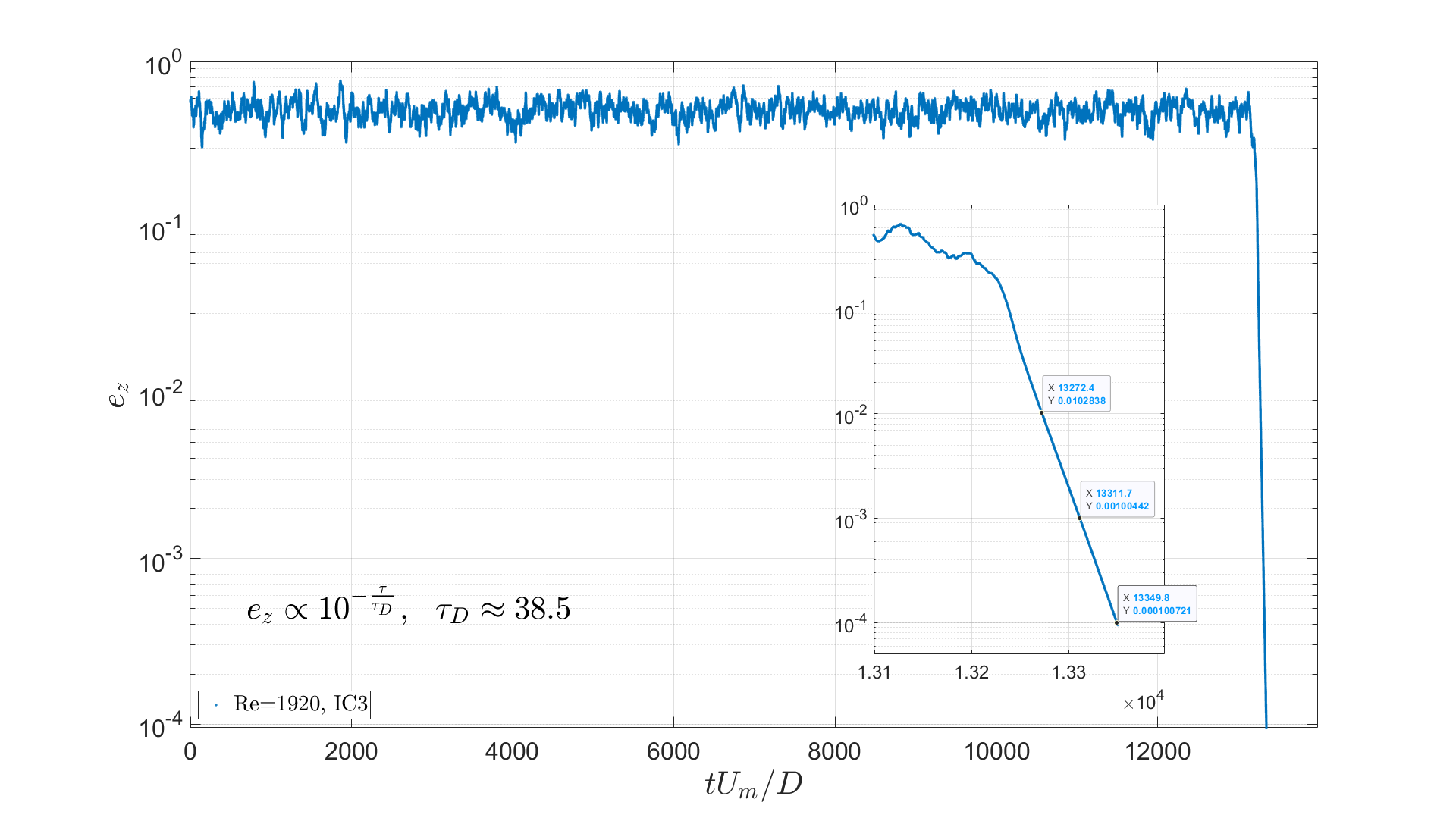}}
}
    \caption{Temporal evolution of the energy of streamwise velocity fluctuations, $e_z$, for different Reynolds numbers;
    in the insets on the right, the last 300--400 $(D/U_m)$ time units, including the relaminarization stage; $\tau=tU_m/D$;
    $\kappa=\tau_D^{-1}$ is the rate of decay. (b) $Re$=1900. $\bullet$ at $\tau$=1089 and $\tau$=1097 represent the final local minimum and maximum before exponential decay, which will be discussed in subsection \ref{subsec: SNpoints}.}
        \label{fig: 1880-1900-1920 ez with tauD}
\end{figure*}

\begin{figure*}
{\centering (a) \hspace{7cm}(b)}
\centerline{
  \hbox{
    \resizebox{70mm}{!}
    %{\includegraphics[width=0.7 \textwidth]{ez-t-250last-Re1880-all}}
    {\includegraphics[width=0.7 \textwidth]{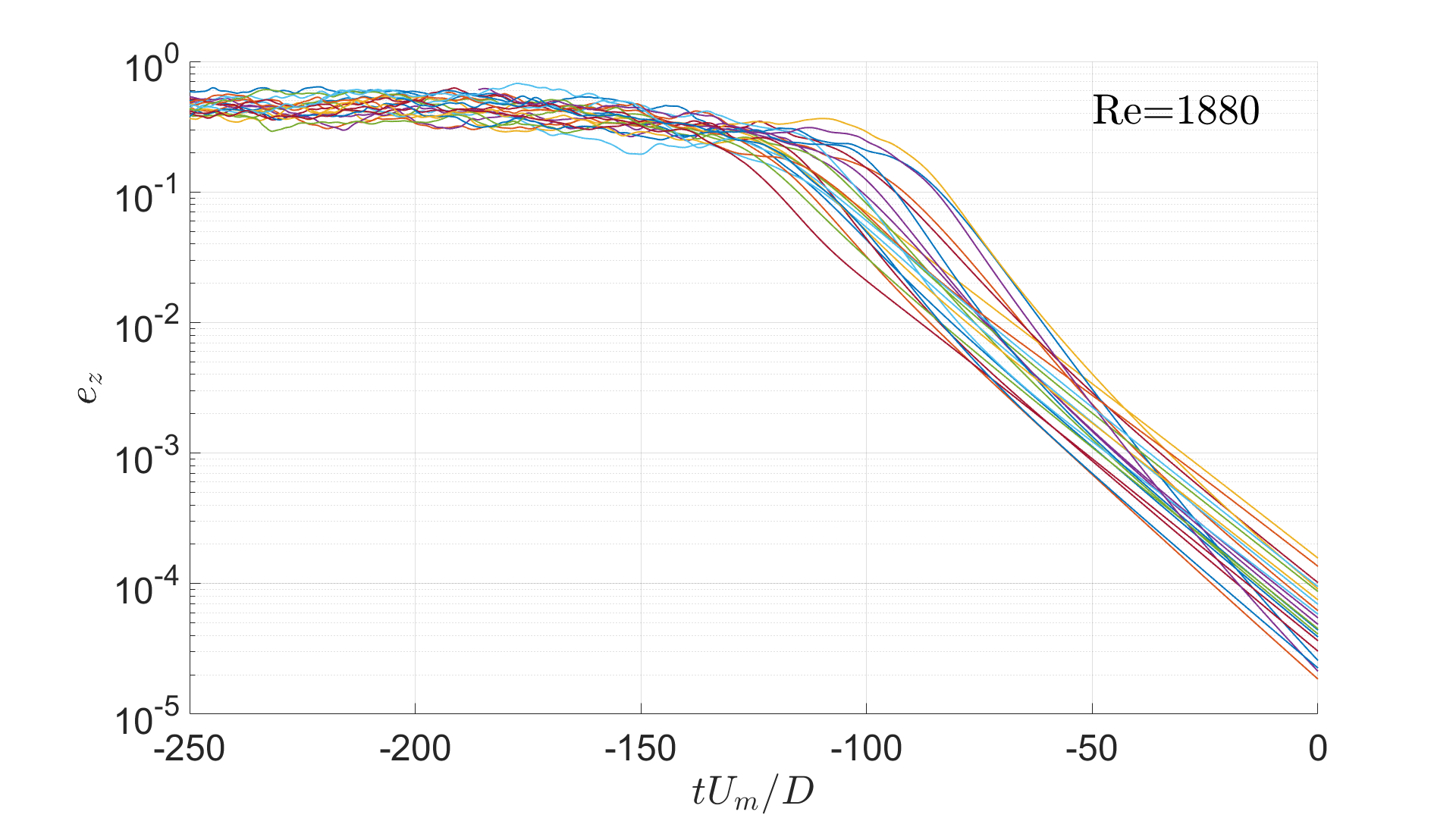}}
    \hspace{1mm}
    \resizebox{70mm}{!}
    %{\includegraphics[width=0.7 \textwidth]{ez-t-250last-Re1900-all}}
    {\includegraphics[width=0.7 \textwidth]{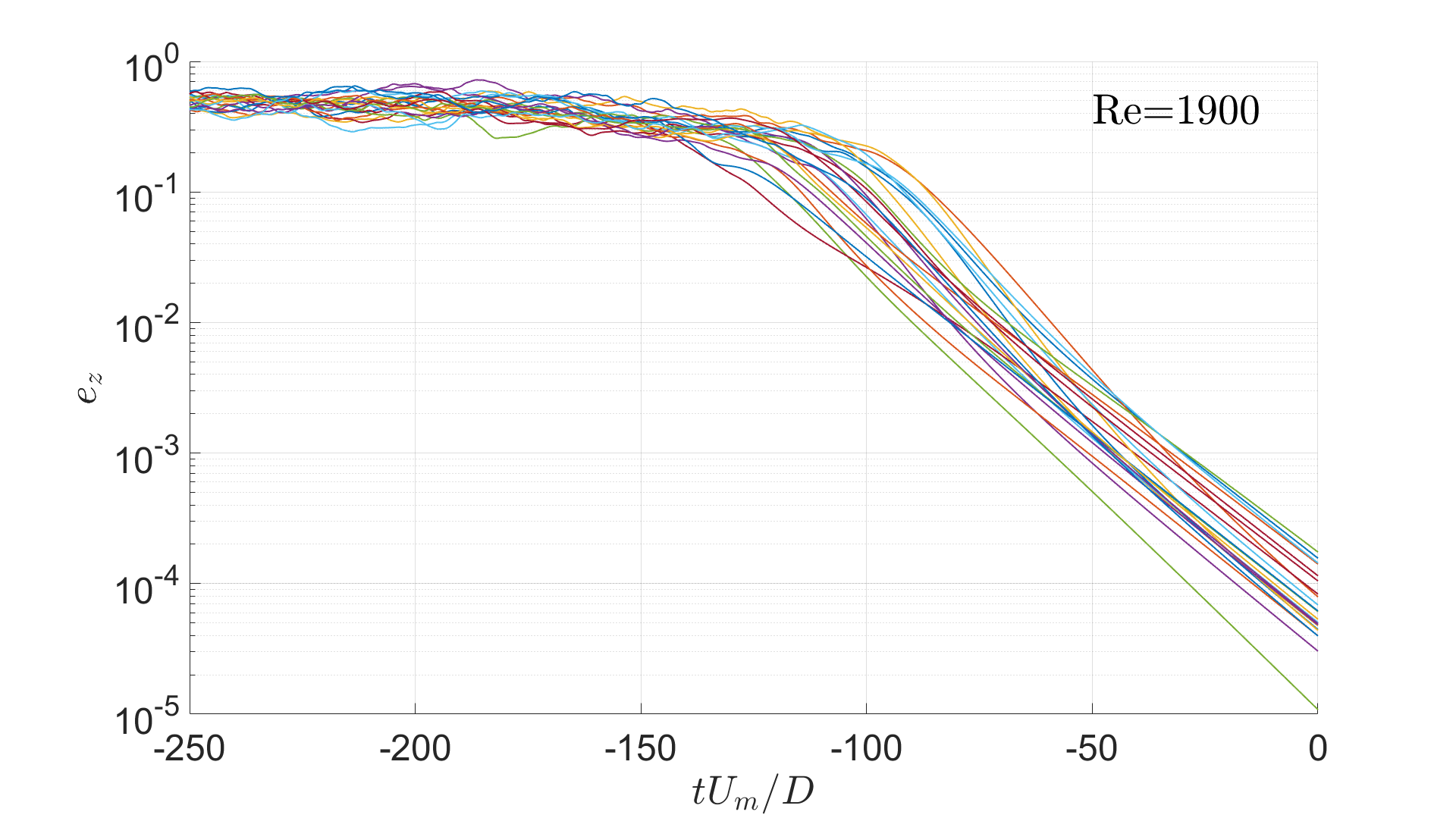}}
    \hspace{1mm}
  }
}
{\centering (c) \hspace{7cm}(d)}
\centerline{
  \hbox{
    \resizebox{70mm}{!}
    %{\includegraphics[width=0.7 \textwidth]{ez-t-250last-Re1920-all}}
    {\includegraphics[width=0.7 \textwidth]{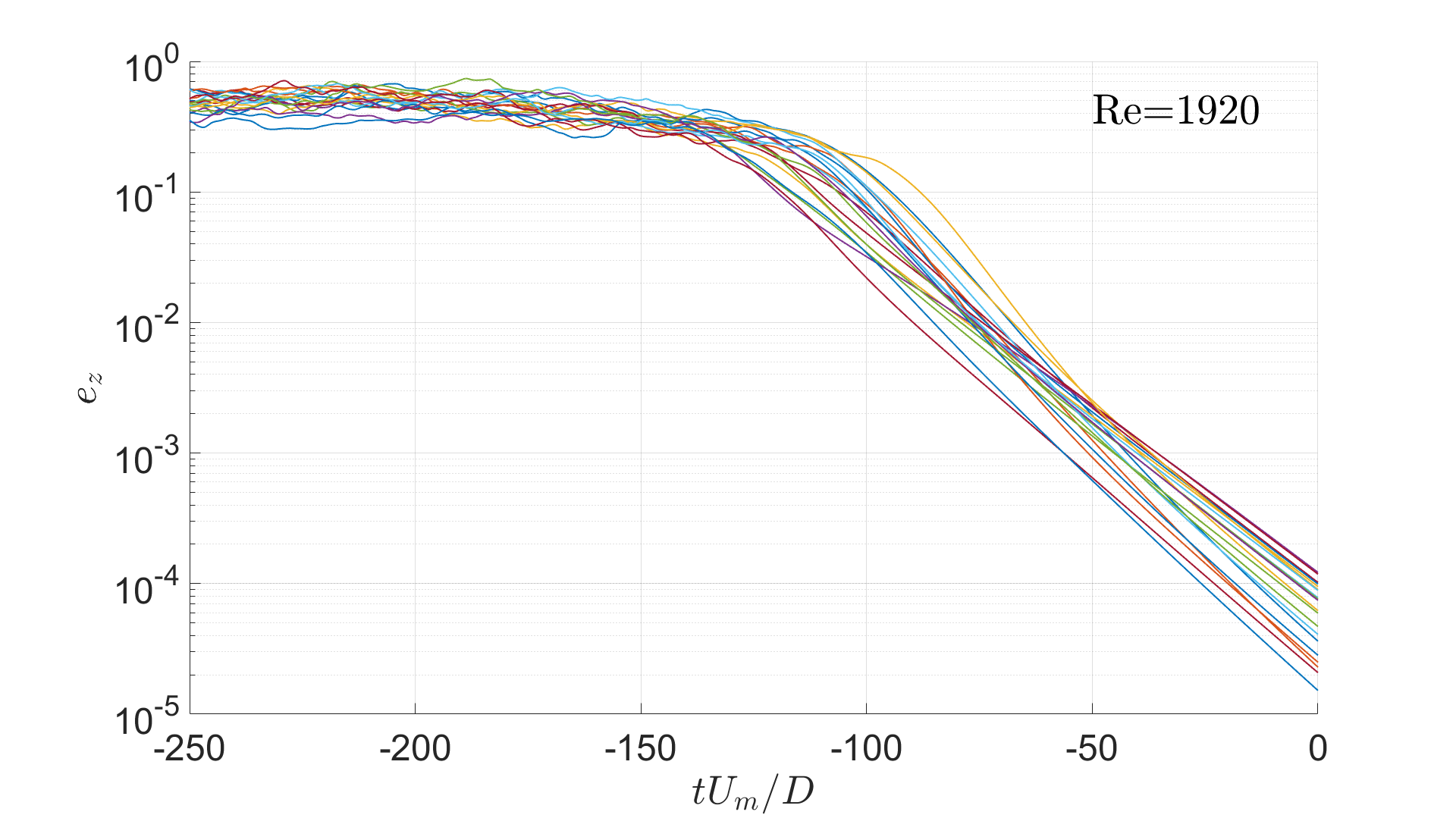}}
    \hspace{1mm}
    \resizebox{70mm}{!}
    %{\includegraphics[width=0.7 \textwidth]{ez-t-250last-Re1880-1900-1920-all}}
    {\includegraphics[width=0.7 \textwidth]{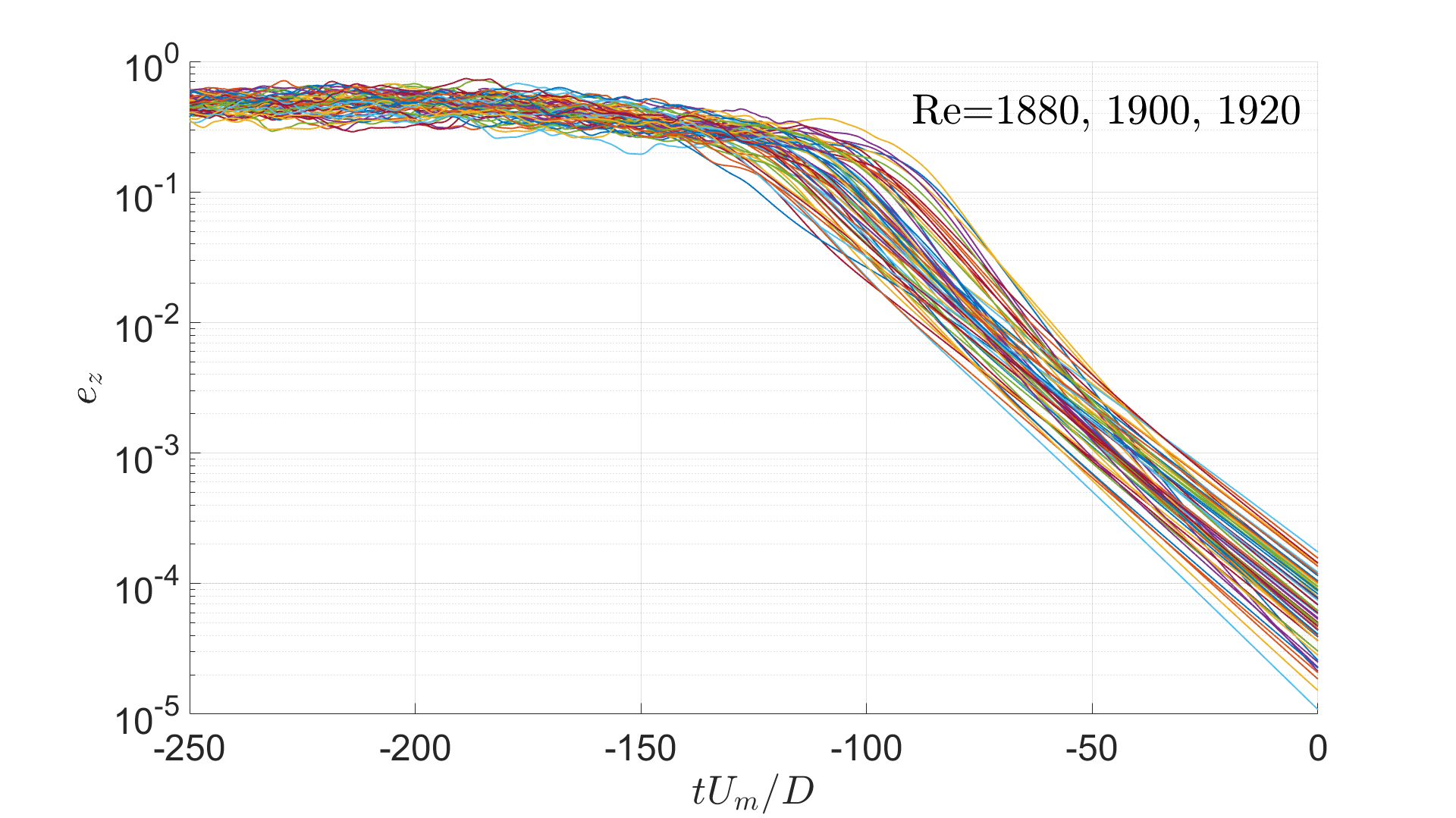}}
    \hspace{1mm}
  }
}
    \caption{
    Time evolution of the energy of longitudinal velocity fluctuations, $e_z$, aligned with the decay time, countdown to complete relaminarization at t=0. (a--c) 22 initial conditions (IC) for each $Re$. (d) countdown before complete relaminarization at t=0, 66 curves.
}
        \label{fig: ez-t-250last_Re1880-1900-1920}
\end{figure*}

The temporal evolutions of the total energy of streamwise velocity fluctuations\footnote{Hereafter, the term 'energy' refers to the total amount of energy: $e_z=\int\int\int u_z^2r\text{d}r\text{d}\theta\text{dz}$}, $e_z$, are shown in figure
\ref{fig: 1880-1900-1920 ez with tauD}. The zoomed-in insets
show the abrupt transition to the relaminarization stage and the exponential decay to laminar flow.
The panels in figure \ref{fig: ez-t-250last_Re1880-1900-1920} display the countdown to full relaminarization, designated as t=0, of the energy of longitudinal velocity fluctuations, $e_z$, for each of the three Reynolds numbers.
The curves in figure
\ref{fig: ez-t-250last_Re1880-1900-1920} imply the exponential decay of streamwise fluctuating energy at the final stage of the relaminarization.
Furthermore, the rate of decay is slightly different for three Reynolds numbers, as evidenced by the collapse of 66 curves in figure \ref{fig: ez-t-250last_Re1880-1900-1920}d.
\comment{The scenario is analogous to the results of DNS of the plane Couette flow\cite{Chantry2014}, in which the streaks' viscous decay is the cause of the exponential decay. In fact, persistent longitudinal large scales are viscously dampened because small eddies decay first and quickly, leaving them without a self-sustaining mechanism when the small eddies transfer energy from the mean shear flow. The distinction is that puffs have a persistent, weakly Reynolds-dependent size that is a characteristic length. Consequently, the presence of a weakly Reynolds-dependent rate of decay is implied\cite{KhanArogetiYakhot2024}.} 

\begin{figure*}
{\centering (a) \hspace{7cm}(b)}
\centerline{
  \hbox{
    \resizebox{70mm}{!}
    %{\includegraphics[width=0.85 \textwidth]{fig4a}}
    {\includegraphics[width=0.85 \textwidth]{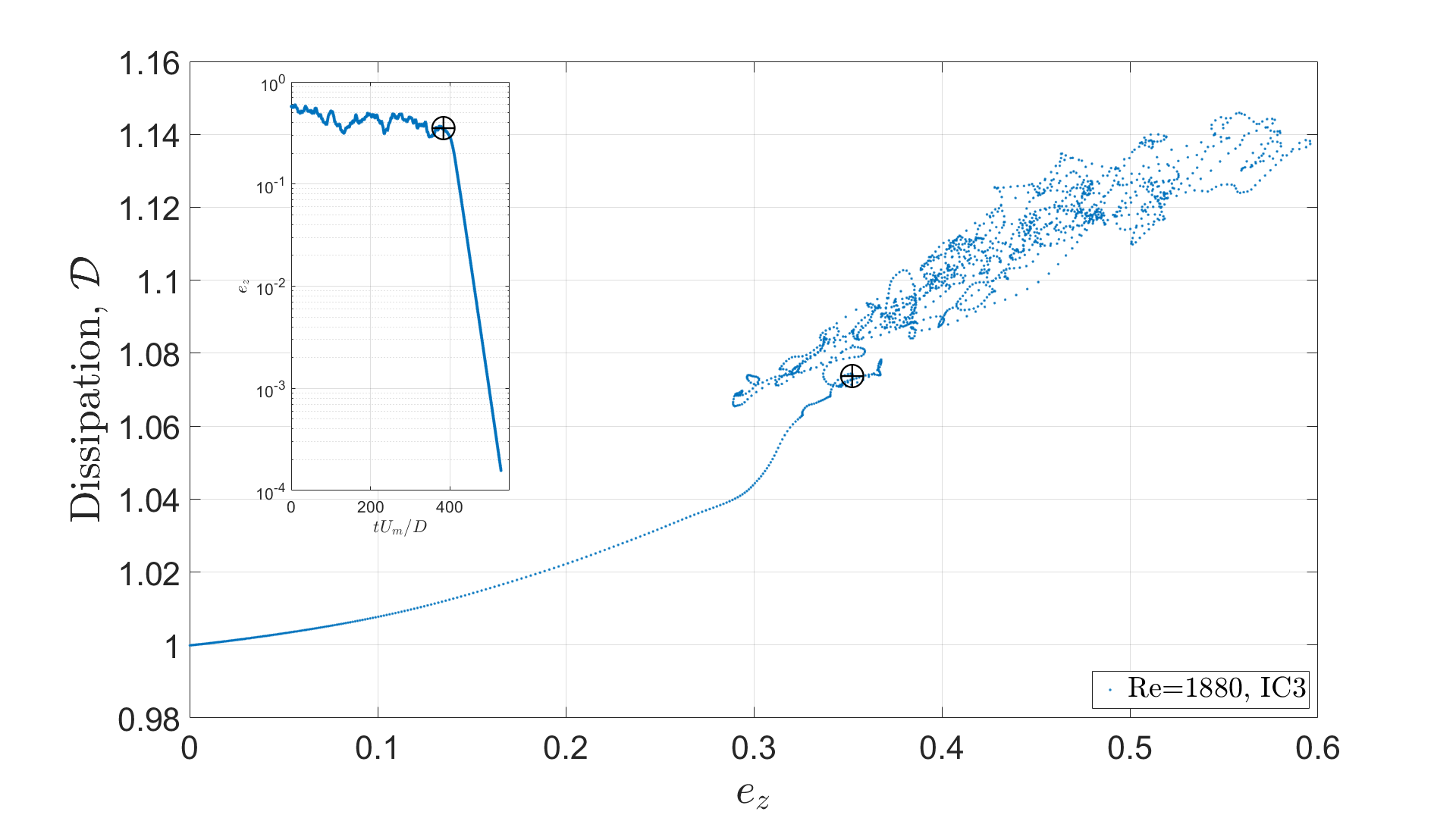}}
    \hspace{1mm}
    \resizebox{70mm}{!}
    %{\includegraphics[width=0.7 \textwidth]{phase-space-ez01-D-Re1880-all}}
    {\includegraphics[width=0.7 \textwidth]{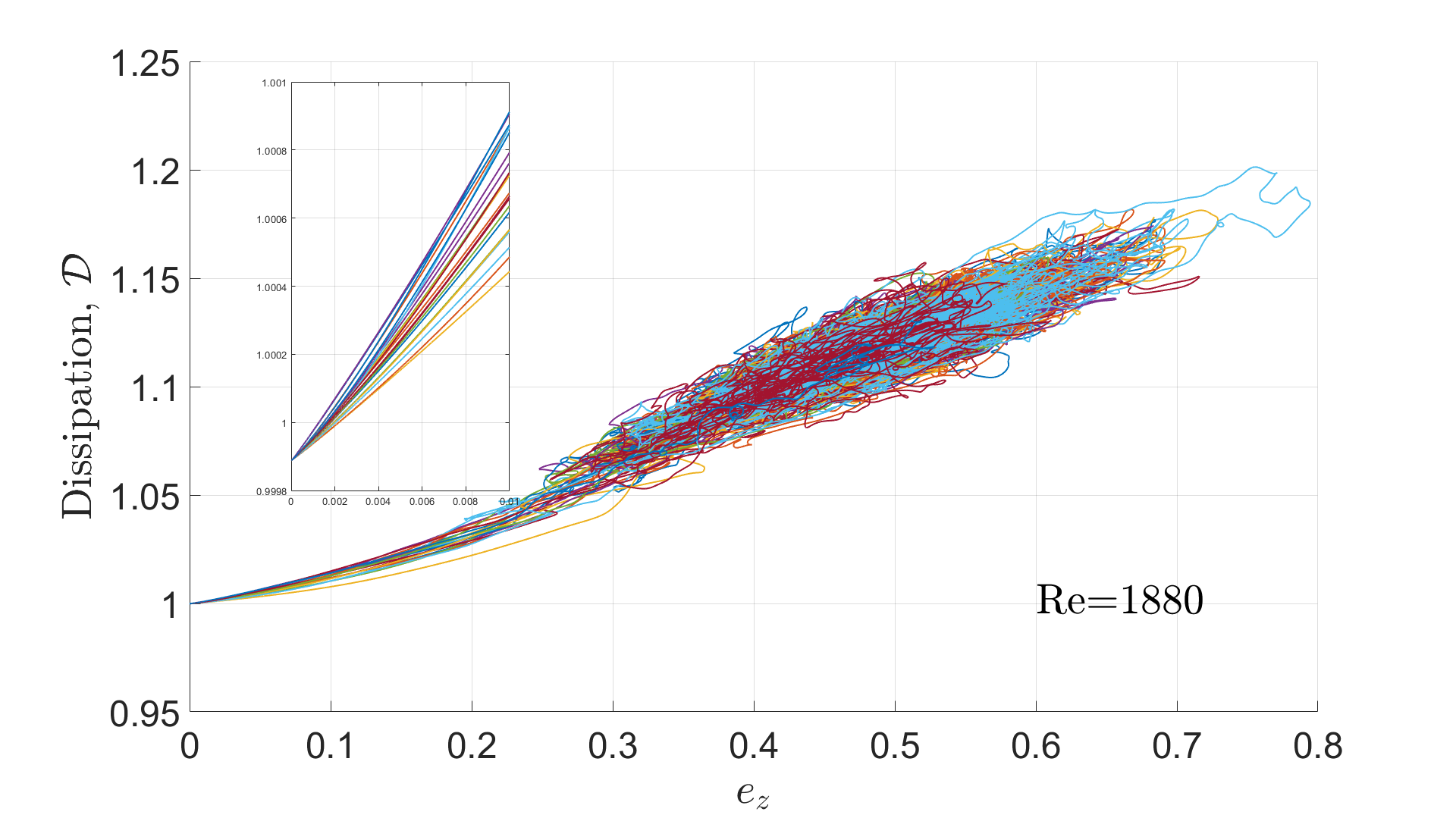}}
    \hspace{1mm}
  }
}
{\centering (c) \hspace{7cm}(d)}
\centerline{
  \hbox{
    \resizebox{70mm}{!}
    %{\includegraphics[width=0.85 \textwidth]{fig4b2}}
    {\includegraphics[width=0.85 \textwidth]{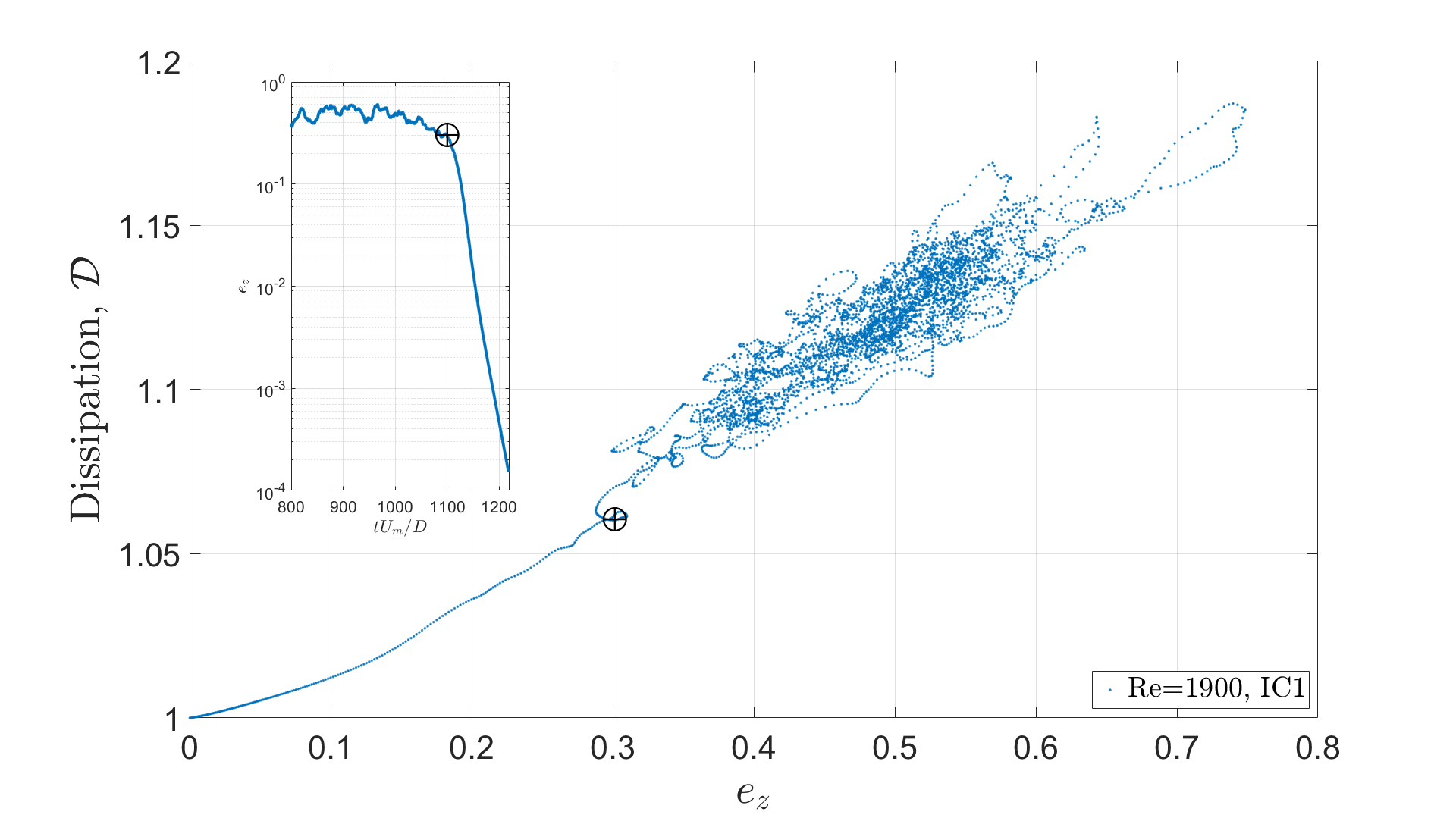}}
    \hspace{1mm}
    \resizebox{70mm}{!}
    %{\includegraphics[width=0.7 \textwidth]{phase-space-ez01-D-Re1900-all}}
    {\includegraphics[width=0.7 \textwidth]{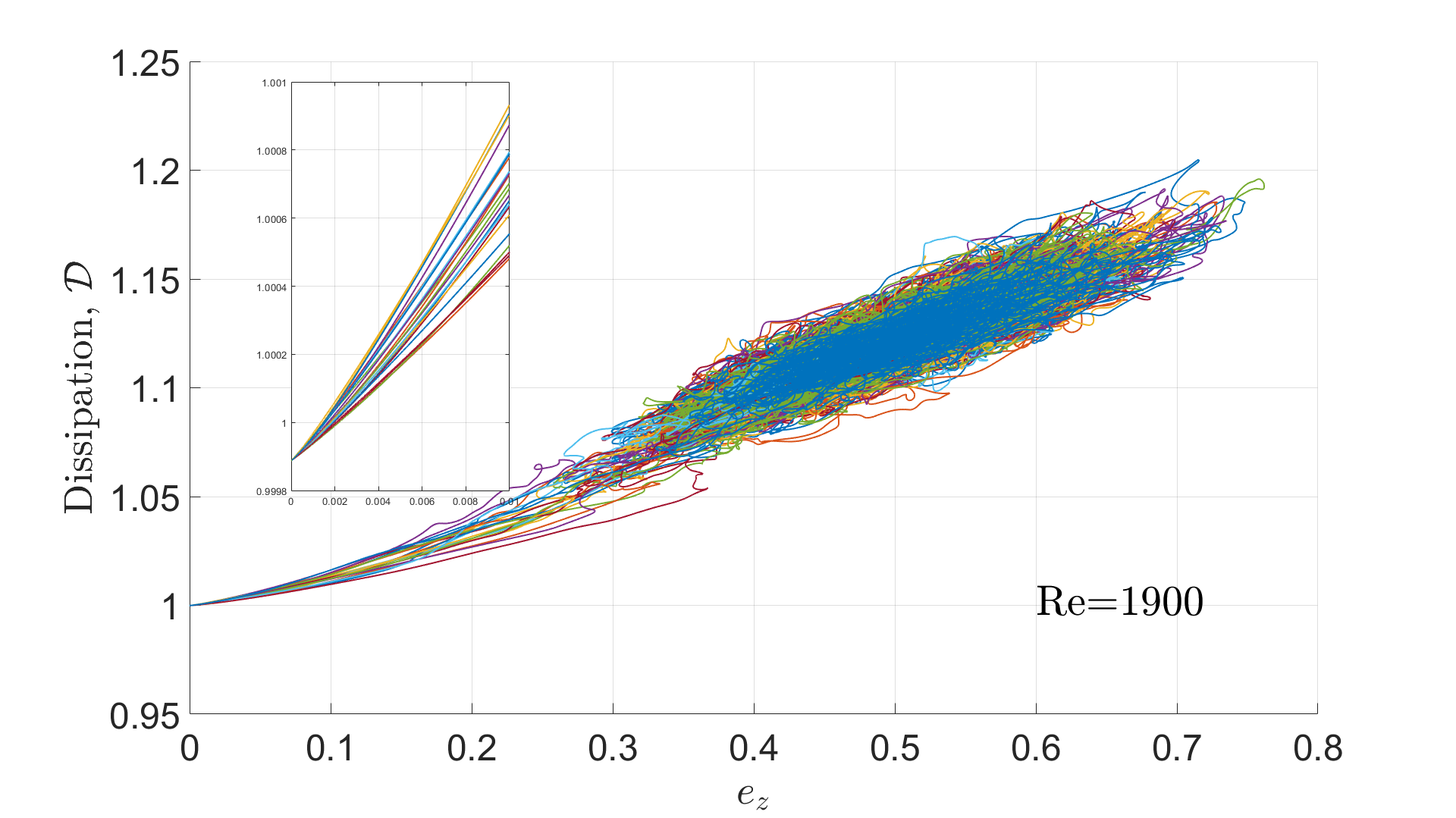}}
    \hspace{1mm}
  }
}
{\centering (e) \hspace{7cm}(f)}
\centerline{
  \hbox{
    \resizebox{70mm}{!}
    %{\includegraphics[width=0.85 \textwidth]{fig4c}}
    {\includegraphics[width=0.85 \textwidth]{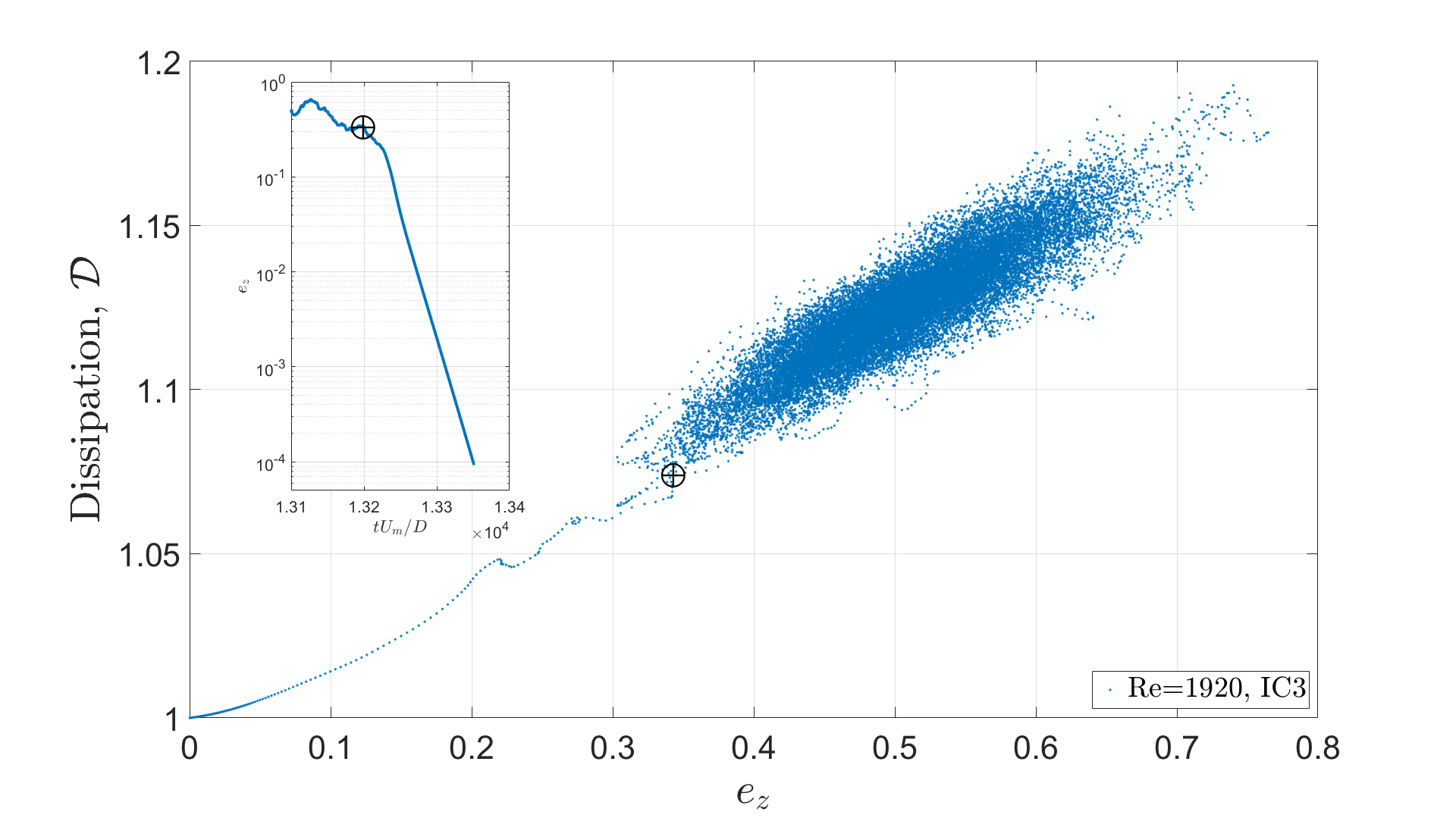}}
    \hspace{1mm}
    \resizebox{70mm}{!}
    %{\includegraphics[width=0.7 \textwidth]{phase-space-ez01-D-Re1920-all}}
    {\includegraphics[width=0.7 \textwidth]{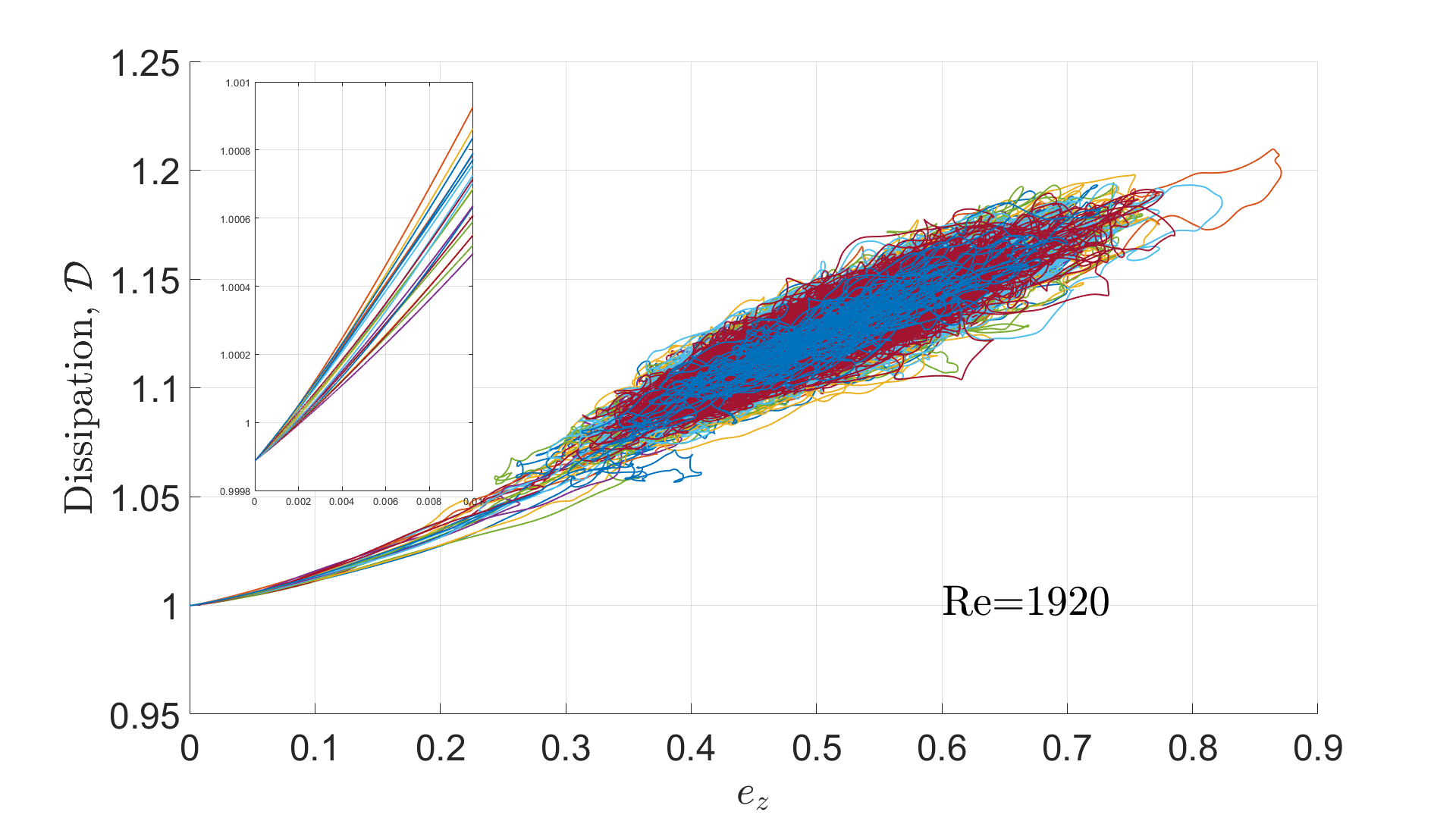}}
    \hspace{1mm}
  }
}
    \caption{
(a, c, e) Phase-space projection of typical chaotic trajectories of the energy of streamwise velocity fluctuations, $e_z$, and the total energy dissipation, $\cal {D}$, computed for different initial conditions (IC) including the relaminarization stage (see figure \ref{fig: 1880-1900-1920 ez with tauD}). (b, d, f) Phase-space projection of $\cal {D}$ vs. $e_z$ trajectories of all 22 initial conditions; in the insets, the last relaminarization stage, $e_z < 0.01$.
}
        \label{fig: ez-vs-D-1880-1900-1920}
\end{figure*}

Figure \ref{fig: ez-vs-D-1880-1900-1920} shows a phase-space projection of typical chaotic trajectories of the energy of
streamwise velocity fluctuations, $e_z$, and the total energy dissipation, $\cal {D}$, computed for different initial conditions
(ICs), including the relaminarization stage. The trajectories exhibit chaotic
motion before
relaminarization. The onset of relaminarization is marked by $\bigoplus$ in the insets. We conventionally consider the onset of relaminarization based on the chaotic trajectories of $\cal {D}$ vs. $e_z$. Specifically, the time beyond which a trajectory shows a gradual decrease in the turbulence energy and dissipation. The insets in figure \ref{fig: ez-vs-D-1880-1900-1920}(b, d, f) demonstrates that the turbulence energy and dissipation are strongly correlated during relaminarization, which is, for instance, the consequence of the viscous exponential decay.

\begin{figure*}
\centerline{
    %{\includegraphics[width=0.8 \textwidth]{tauD-all-Re+inset}} %\Papers\TurboDecay\mCodes2\fit_tauDm_Q3m_all_Re_with_inset.m
    {\includegraphics[width=0.8 \textwidth]{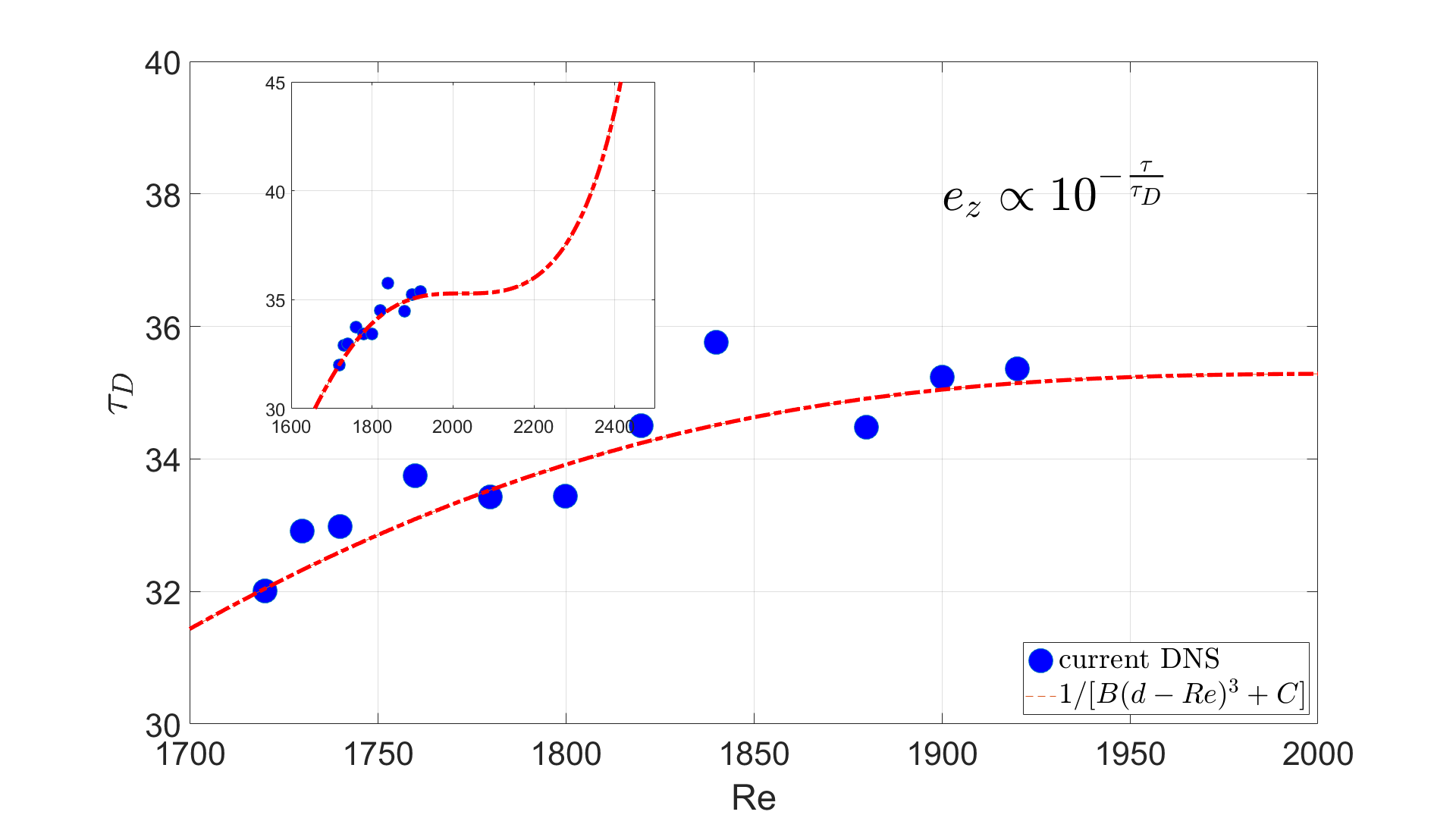}} %\Papers\TurboDecay\mCodes2\fit_tauDm_Q3m_all_Re_with_inset.m
}
    \caption{The rate of decay, $\tau_D$, of the energy of streamwise velocity component fluctuations,
    $e_z \propto 10^{-\tau/\tau_D}$.
    }
        \label{fig: tauD_allRe_1770-1920}
\end{figure*}

Figure \ref{fig: tauD_allRe_1770-1920} shows the decay rate, $\tau_D$, of the streamwise velocity fluctuation energy,
$e_z \propto 10^{-\tau/\tau_D}$. In this form, a decay rate of $\tau_D \approx 33$ implies that as
a turbulent puff moves downstream a distance $L \approx 120D$ at a speed of $0.9U_m$, the turbulent energy decreases by
four orders of magnitude. The symbols represent the mean values obtained from 22 different initial
conditions
\comment{
\footnote{Among the 22 values of $\tau_D$ obtained under different initial conditions, the lowest and highest values were excluded from the calculation of the average.}
}
for each Reynolds number. The dashed curve is the fitting expression of
$\kappa=\tau_D^{-1}=B(d-Re)^3+C,~B=10^{-10},~d=2024,~C=0.028$. The inset is added to show the correct trend of the
approximation with increasing the Reynolds number. Note that the Sreenivasan's cubic approximation\cite{NarasimhaSreeni1979}
(without adding the constant C) seems to work well for low Reynolds numbers on the descending branch of the approximation. Two branches of the cubic approximation curve $\kappa=\kappa(Re)$ correctly reflect the expected trend of the puff lifetime.
Indeed, for $Re<d$, $\kappa$ increases, which means shortening the lifetime; for $Re>d$, $\kappa$ decreases, which means
the lifetime increases.

\begin{figure*}
(a)\hspace{4.5cm}(b)\hspace{4.5cm}(c)
\centerline{
  \hbox{
    \resizebox{45mm}{!}
    %{\includegraphics[width=3.5cm, height=1.5cm]{fig3a}}
    {\includegraphics[width=3.5cm, height=1.5cm]{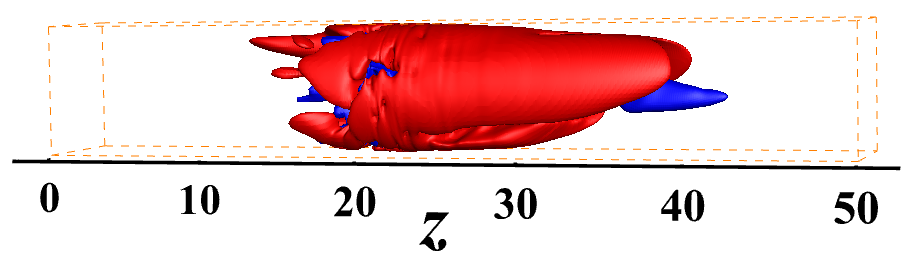}}
    \hspace{1mm}
    \resizebox{45mm}{!}
    %{\includegraphics[width=3.5cm, height=1.5cm]{fig3b}}
    {\includegraphics[width=3.5cm, height=1.5cm]{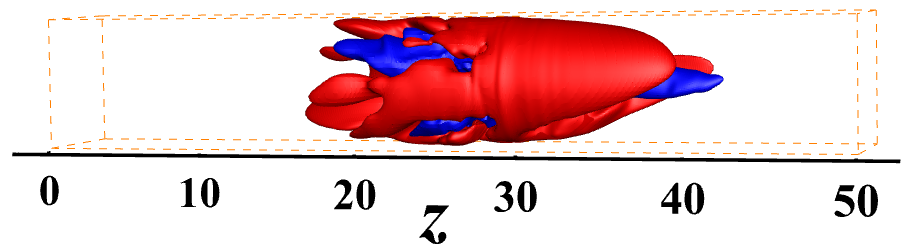}}
    \hspace{1mm}
    \resizebox{45mm}{!}
    %{\includegraphics[width=3.5cm, height=1.5cm]{fig3c}}
    {\includegraphics[width=3.5cm, height=1.5cm]{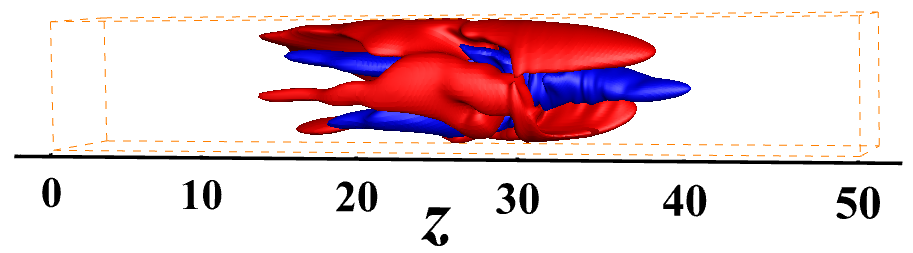}}
  }
}
% (a)\hspace{5.5cm}(b)\hspace{5.5cm}(c)
    \caption{
$Re=1900$. Iso-surfaces of the streamwise velocity fluctuations $u_z$ are shown to illustrate the size of a turbulent puff relative to the computational domain during the onset of decay stage (figure \ref{fig: 1880-1900-1920 ez with tauD}b); (a) $tU_m/D=1000$, (b) $tU_m/D=1050$ and (c) $tU_m/D=1087$.
}
        \label{fig: 1900_uz-4D3D4D-dif_times3}
\end{figure*}

In conclusion, we suggest a straightforward estimate of the rate of decay.
Defining spatially averaged dissipation and introducing characteristic length ($l_c$) and velocity ($u_c$) during the relaminarization stage yield:
$\frac{de_{z}}{dt}=-{\cal D}\propto -\frac{e_z}{t_c}=-\frac{e_z}{l_c/u_c}$. Solution to this equation reads
\footnote{Recasting the solution (\ref{eq: sol_dedt=-ke}), we used assumptions regarding the characteristic velocity
($u_c$) and length ($l_c$).}:
\begin{align}\label{eq: sol_dedt=-ke}
  e_z \propto \exp \left(-\frac{tu_c}{l_c}\right)=\exp \left(-\frac{tU_p}{\alpha L_p}\right)
  =\exp \left[-\frac{t(0.9U_m)}{\alpha L_p}\right]\nonumber\\
  =\exp \left[-\left(\frac{0.9}{\alpha}\right)\cdot\frac{tU_m}{(nD)}\right]
  =10^{\left[-\left(\frac{0.9\log(\text{e})}{\alpha n}\right)\cdot\frac{tU_m}{D}\right]}
  =10^{-\frac{\tau}{\tau_D^*}},\nonumber\\
  \text{~where~}\tau_D^*=\frac{\alpha n}{0.9\log(\text{e})}.
\end{align}
In this equation, $L_p$ and $U_p$ stand for the puff's length and velocity, respectively; $L_p = nD$, $n$ and $\alpha$
are  scaling coefficients. In figure \ref{fig: 1900_uz-4D3D4D-dif_times3}, the iso-surfaces of streamwise velocity
fluctuations ($u_z$) are depicted for three different
time instances to illustrate the size of a turbulent puff at the onset of the decay stage
(figure \ref{fig: 1880-1900-1920 ez with tauD}b). As for the estimate of $n$, figure
\ref{fig: 1900_uz-4D3D4D-dif_times3} suggests $n\approx 25$, which means $\tau_D^*=64\alpha$. The scaling coefficient $\alpha$
accounts for the reduction in puff's size (``shrinking'') during the decay stage. Thus, a rough estimate of $\alpha=0.5$ seems
reasonable\cite{Peixinho2006}, and we obtain $\tau_D^*=32$, which is consistent with the DNS results depicted in figure
\ref{fig: tauD_allRe_1770-1920}.

\subsection{Inflection points}\label{subsec: IPoints}
Figure \ref{fig: Uc(z)-Uz(r,th)_z=-8-6-4_Re1900_t1000} (upper panel) shows the centerline velocity along the pipe. Upstream
of the turbulent puff ($z<-7D$), the flow is nearly laminar with $U_c\approx 1$. The sharp changes observed between $z=-3D$
and $z=-D$ at the three selected times indicate the presence of a sharp interface between the laminar and turbulent regions
near the upstream edge of the puff. In particular, at time $t=1000$ the interface is located at $z=-2D$, which corresponds
to $z=0$ in figure 1 of Ref.~\onlinecite{Hof2010}.
In figure \ref{fig: Uc(z)-Uz(r,th)_z=-8-6-4_Re1900_t1000} (lower panels), the instantaneous streamwise velocity profiles along
the puff are displayed at various azimuthal locations. At $z = -8D$, the laminar parabolic velocity profile remains nearly
undisturbed; however, at $z = -4D$, the onset of turbulence is evident from a slight decrease in centerline velocity (top
panel) and the appearance of an inflection point in the velocity profile (bottom panels, $z = -4D$). Here, the virtually
undistorted, fast-moving laminar flow crosses the laminar-turbulent interface, maintaining a high mean flow shear, which is
essential for the energy transfer to the eddies for sustaining turbulence. At the same time, the inflection points in
the longitudinal velocity profile are the source of instability for triggering turbulence\cite{Hof2010}.
A short distance of two diameters downstream (figure \ref{fig: Uz_dif_z_Re1900_t1000}), there is a significant buildup
of velocity profiles with noticeable inflection points. Furthermore, this occurs almost without distorting the parabolic profile
at the core but rather due to acceleration near the wall (see panels at $z=-3D$ and $z=-2D$ in
figure \ref{fig: Uz_dif_z_Re1900_t1000}). Between the locations z=-2D and z=0 (see figure \ref{fig: Uz_dif_z_Re1900_t1000}),
the mean profile is significantly distorted. The turbulent production is very large, resulting in the maximum in-plane energy
($e_\bot=e_r+e_\theta$) being reached at $z=0$.
The strength of inflection point was estimated using the procedure outlined in the Online Supporting Material of Ref.
\onlinecite{Hof2010}, which was based on the change in velocity profile curvature. This procedure of summing the positive values
of the second derivatives $U_z^{\prime \prime}(r)$ was implemented over the interval $0.4\leq r/R \leq 0.8$, where the inflection
points are most pronounced. As shown in figure \ref{fig: Uc(z)-Uz(r,th)_z=-8-6-4_Re1900_t1000} (upper panel), the strongest
inflection point index (IPindex) is detected in the interval
$-1\leq z\leq 0$. The centerline velocity ($U_c$) reaches its minimum value at $z=0$. Downstream (in the frame of reference
moving with the puff), the region $z > 0$ indicates a slow recovery to laminar flow. The panels at $z=0$ and $z=2D$ in
figure \ref{fig: Uz_dif_z_Re1900_t1000} show the onset of recovery linked to a flattening of the core velocity profile.
This causes a decrease in energy transfer to turbulent eddies, destroying the self-sustaining process of turbulence and
ultimately resulting in laminarization\cite{Hof2010}. In the interval $0 < z < 2D$, our current and previous
findings demonstrate a ``turbulent flash'' similar to a fully-developed turbulent flow. Along with a flattening velocity
profile, the near-wall energy distribution of longitudinal intensities at $y_+ < 15$ is similar to that in a fully developed
turbulent flow in a pipe and reaches a peak at $y_+ = 15$ ($y_+$ is the distance from the wall, expressed in wall
units)\cite{Yakhot2019}.
\begin{figure*}
\centerline{
  %{\includegraphics[width=1.0 \textwidth]{../../Figures/SNB_Uz/SNB_Uz_diff_z/Uc-Re1900IC1-800-900-1000_with_IPindex10_t1000}}
  \includegraphics[width=0.8 \textwidth]{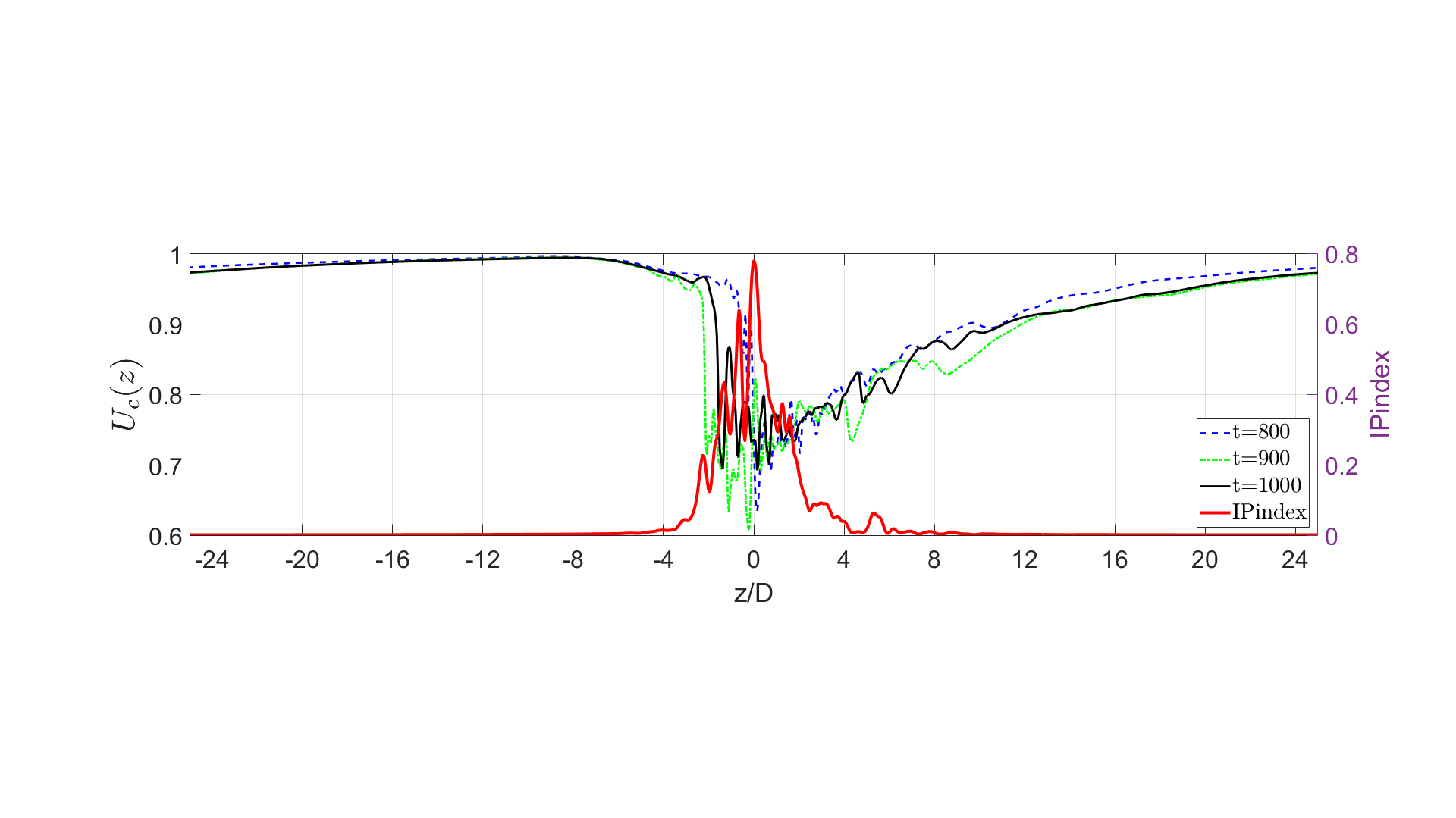}

}
\centerline{
  \hbox{
    \resizebox{55mm}{!}
    %{\includegraphics[width=0.7 \textwidth]{../../Figures/SNB_Uz/SNB_Uz_diff_z/Uz-at_z-8d_Re1900_t1000b}}
    {\includegraphics[width=0.85 \textwidth]{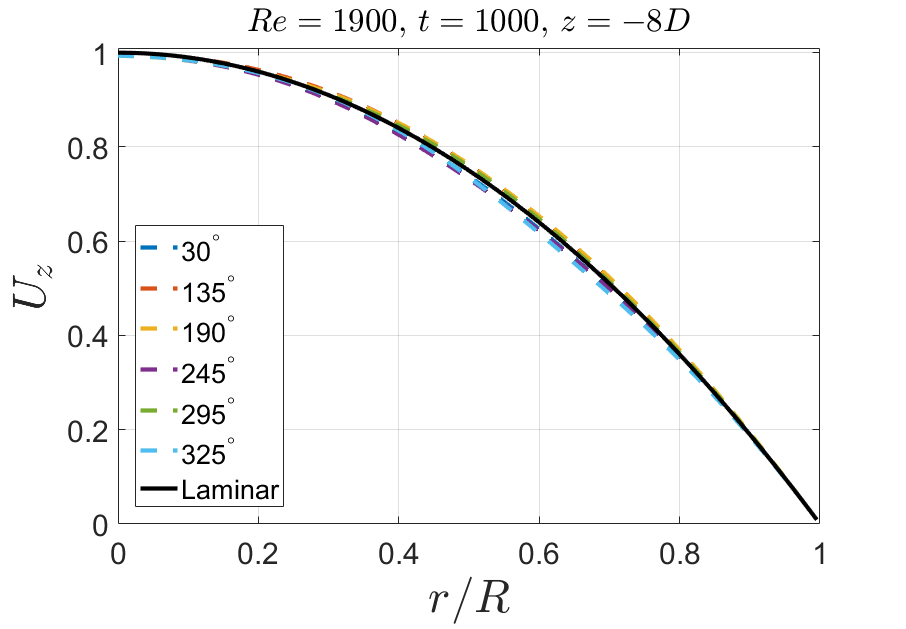}}
    \hspace{1mm}
    \resizebox{55mm}{!}
    %{\includegraphics[width=0.7 \textwidth]{../../Figures/SNB_Uz/SNB_Uz_diff_z/Uz-at_z-6d_Re1900_t1000b}}
    {\includegraphics[width=0.85 \textwidth]{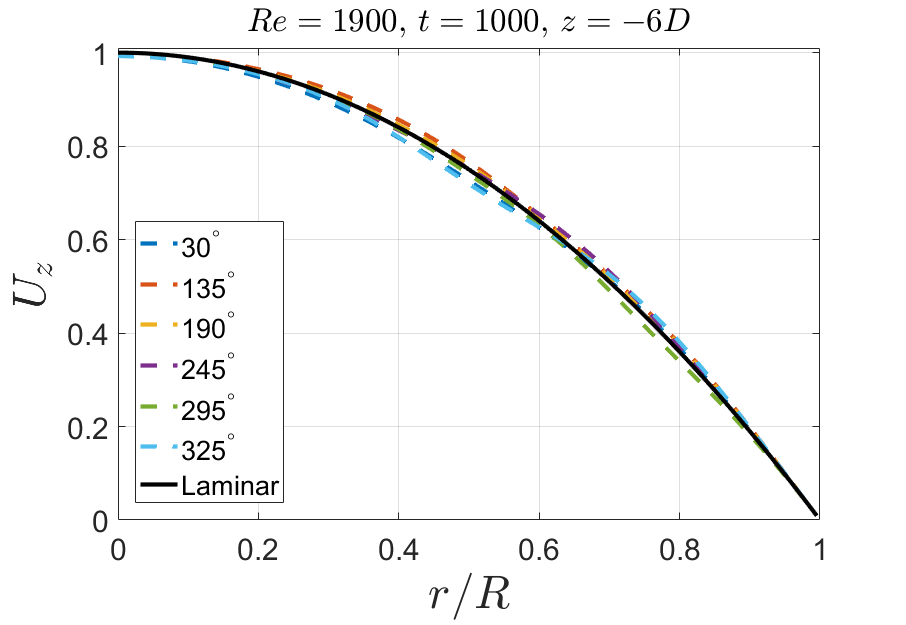}}
    \hspace{1mm}
    \resizebox{55mm}{!}
    %{\includegraphics[width=0.7 \textwidth]{../../Figures/SNB_Uz/SNB_Uz_diff_z/Uz-at_z-4d_Re1900_t1000b}}
    {\includegraphics[width=0.85 \textwidth]{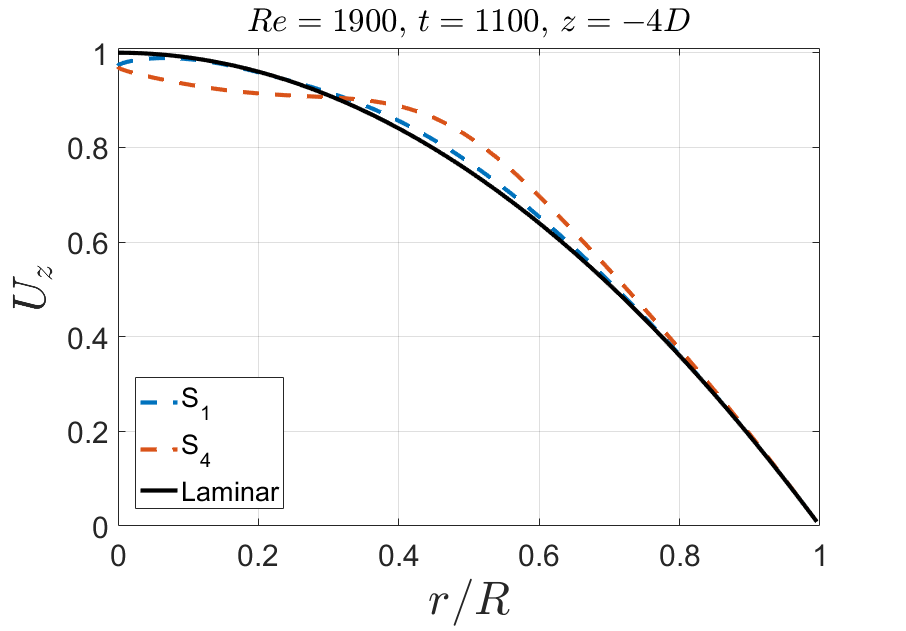}}
  }
}
    \caption{
$Re=1900$, IC1. (top panel, the flow direction from left to right) Centerline velocity in a pipe at various time instances during the self-sustaining, statistically steady-state stage (see figure \ref{fig: 1880-1900-1920 ez with tauD}b), and the strength index of the inflection point\cite{Hof2010} at t=1000. (bottom panels) Instantaneous streamwise velocity profiles at different azimuthal locations at different locations along a puff; $z=0$ corresponds to the location, where the energy of transverse (turbulent) motion is maximal, while $z=-2D$ corresponds to $z=0$ in Fig.1 in  Ref.~\onlinecite{Hof2010} (Hof Science2010); $t$ in $D/U_m$ units.
}
        \label{fig: Uc(z)-Uz(r,th)_z=-8-6-4_Re1900_t1000}
\end{figure*}
\begin{figure*}
\centerline{
  \hbox{
    \resizebox{65mm}{!}
    %{\includegraphics[width=0.7 \textwidth]{../../Figures/SNB_Uz/SNB_Uz_diff_z/Uz-at_z-3d_Re1900_t1000b}}
    {\includegraphics[width=0.7 \textwidth]{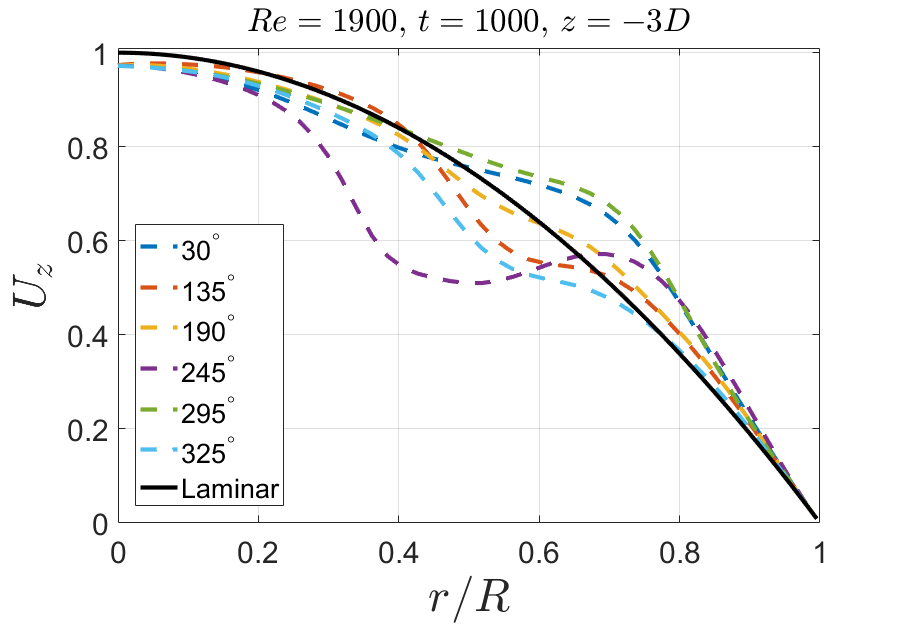}}
    \hspace{1mm}
    \resizebox{65mm}{!}
    %{\includegraphics[width=0.7 \textwidth]{../../Figures/SNB_Uz/SNB_Uz_diff_z/Uz-at_z-0d_Re1900_t1000b}}
    {\includegraphics[width=0.7 \textwidth]{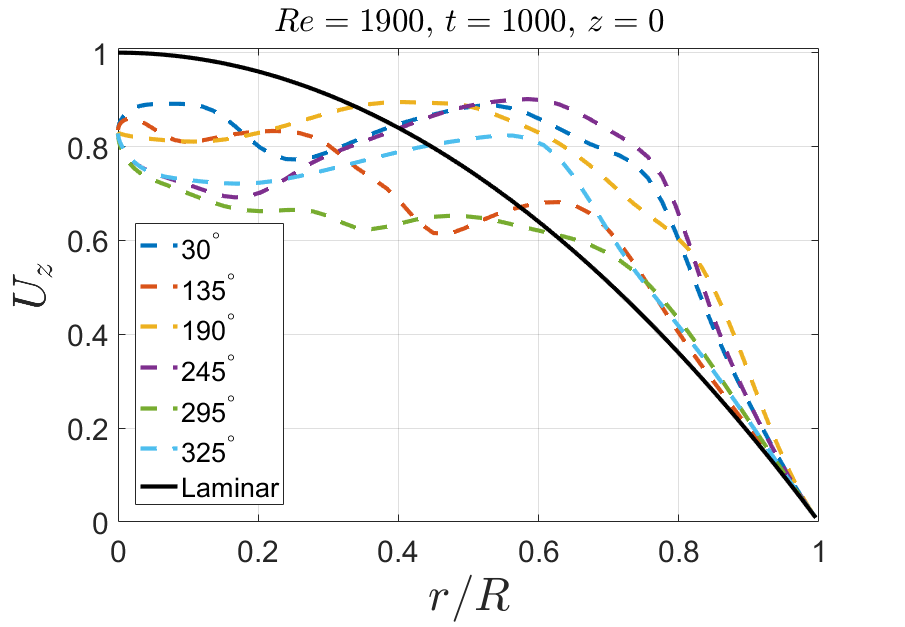}}
    \hspace{1mm}
  }
}
\centerline{
  \hbox{
    \resizebox{65mm}{!}
    %{\includegraphics[width=0.7 \textwidth]{../../Figures/SNB_Uz/SNB_Uz_diff_z/Uz-at_z-2d_Re1900_t1000b}}
    {\includegraphics[width=0.7 \textwidth]{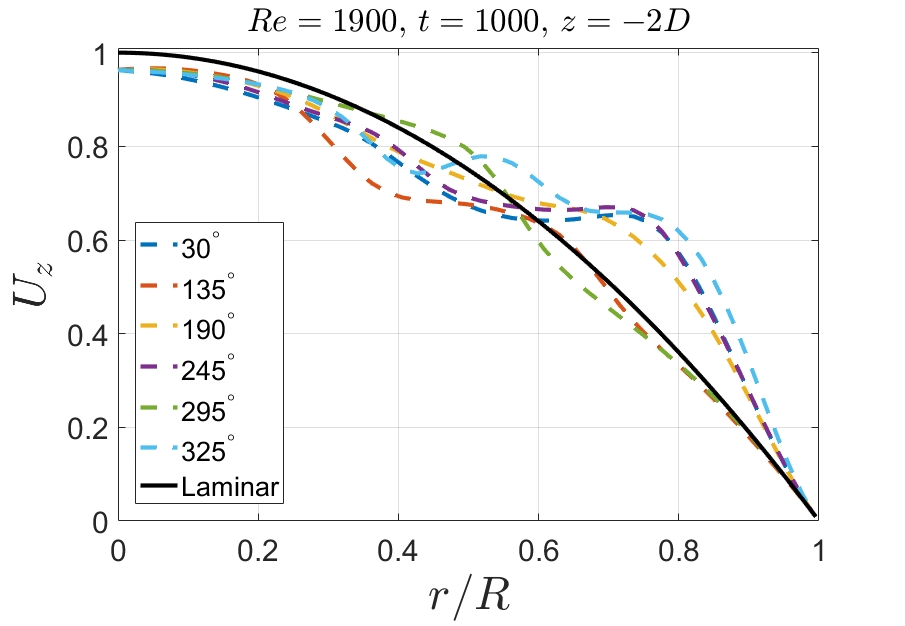}}
    \hspace{1mm}
    \resizebox{65mm}{!}
    %{\includegraphics[width=0.7 \textwidth]{../../Figures/SNB_Uz/SNB_Uz_diff_z/Uz-at_z+2D_Re1900_t1000b}}
    {\includegraphics[width=0.7 \textwidth]{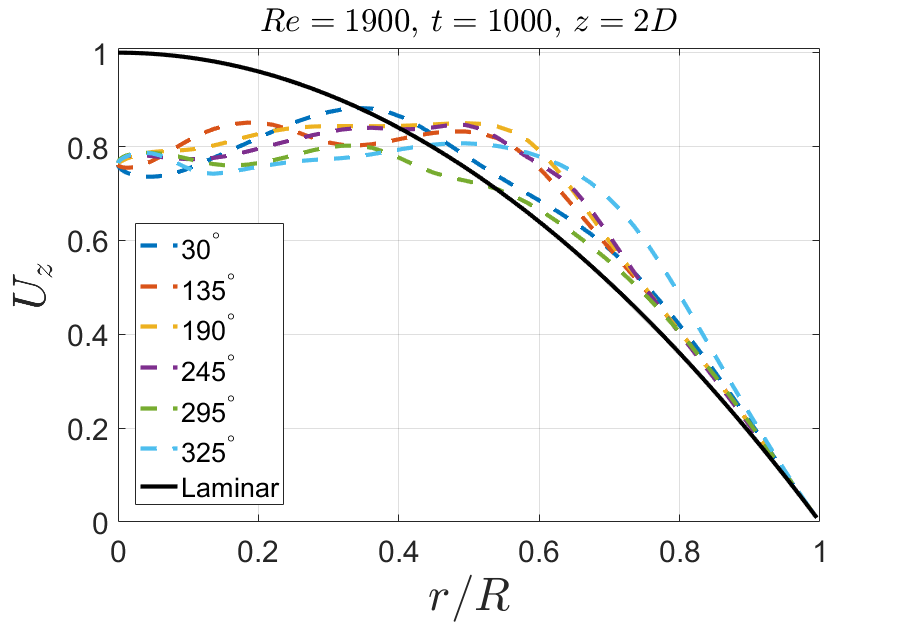}}
    \hspace{1mm}
  }
}
\centerline{
  \hbox{
    \resizebox{65mm}{!}
    %{\includegraphics[width=0.7 \textwidth]{../../Figures/SNB_Uz/SNB_Uz_diff_z/Uz-at_z-1D_Re1900_t1000b}}
    {\includegraphics[width=0.7 \textwidth]{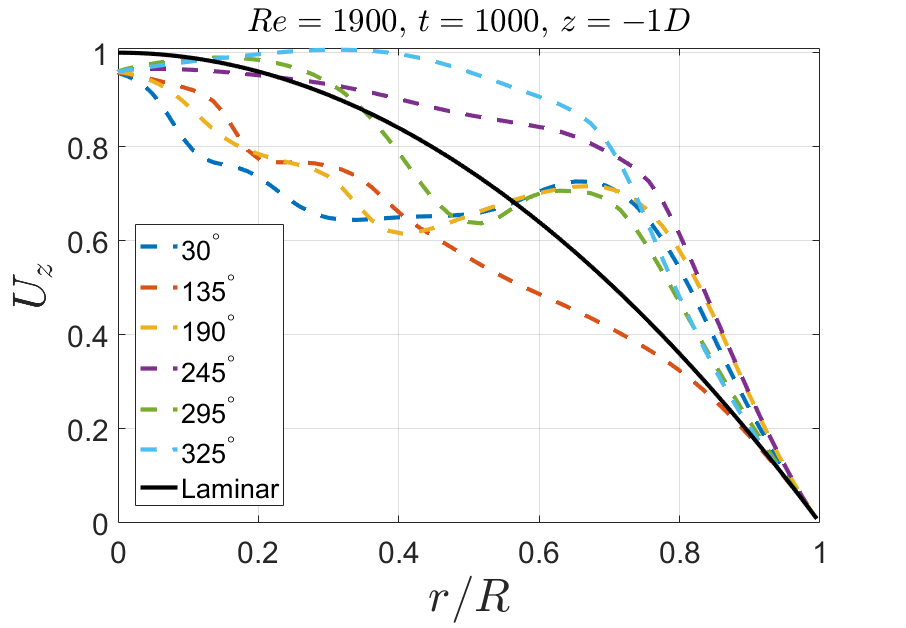}}
    \hspace{1mm}
    \resizebox{65mm}{!}
    %{\includegraphics[width=0.7 \textwidth]{../../Figures/SNB_Uz/SNB_Uz_diff_z/Uz-at_z+4D_Re1900_t1000b}}
    {\includegraphics[width=0.7 \textwidth]{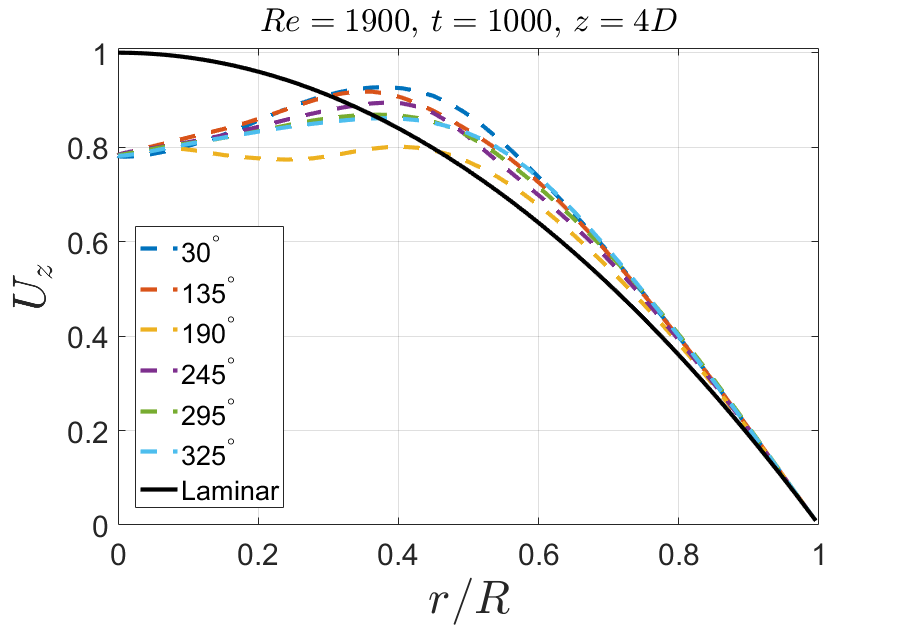}}
    \hspace{1mm}
  }
}
    \caption{
$Re=1900$, IC1. (upper panel, the flow direction from left to right) Centerline velocity along a pipe for three time points in the self-sustaining, statistical
steady-state stage (figure \ref{fig: 1880-1900-1920 ez with tauD}b). (two columns) Instantaneous streamwise velocity profiles at different locations along a puff; $z=0$ corresponds to the location, where the energy of transverse (turbulent) motion is maximal, and $z=-2D$ corresponds to $z=0$ in Fig.1 in  Ref.~\onlinecite{Hof2010} (Hof Science2010); $t$ in $D/U_m$ units.
}
        \label{fig: Uz_dif_z_Re1900_t1000}
\end{figure*}
\subsection{Saddle and nodal points }\label{subsec: SNpoints}
Figures \ref{fig: Uc(z)-Uz(r,th)_z=-8-6-4_Re1900_t1000} and \ref{fig: Uz_dif_z_Re1900_t1000} display the results which are
consistent with those reported in Ref. \onlinecite{Hof2010}, especially regarding the locations of the strongest inflection point
in velocity profile, the sudden drop in centerline velocity, and the maximum turbulent kinetic energy. The inflection point in
the velocity profile means that the expression $(u_z-U_z)$ changes sign, that is, the axial movement relative to the average
velocity is in the opposite direction. Inflection points are known to be a necessary condition for instability which manifests as
formation of vortical regions. The local motion in opposite directions relative to the mean velocity is implicit in the topology
of the inflection point in the mean axial velocity profile. A mechanism for supporting the inflection points was identified
by examining the vorticity transport ($P_\omega$), which is defined as the in-plane average of the product of the axial vorticity
magnitude and the axial motion relative to the mean velocity\cite{Hof2010}:
$P_\omega=<|\partial u_\theta/\partial r-\partial u_r/\partial \theta|(u_z-U_z)>$. The inflection points lead to instability, which sustains turbulence onset and restores vorticity. Even weak inflection points cause instability, which subsequently induces turbulence (Online Supporting Material of Ref.
\onlinecite{Hof2010}). In subsection \ref{subsec: IPoints}, the location $z=-4D$ is designated as the onset of turbulence
based on both the small value of the IPindex and the weakly expressed inflection point
(figure \ref {fig: Uc(z)-Uz(r,th)_z=-8-6-4_Re1900_t1000}). Thus, the flow regime near $z=-4D$ is intermittently
laminar-turbulent. In this regard, saddle points are inherent topological features of the velocity field that arise when
distinct flow regions interact, that is, the flow is intermittent\cite{BissetHuntRogers2002}.
In Ref. \onlinecite{BissetHuntRogers2002}, studies of the turbulent wake behind a flat plate have identified saddle points in the
flow structure near the moving interface between fully turbulent and non-turbulent states. This was made possible by
the analyzing the in-plane (sectional) streamlines
\footnote{Sectional streamlines are lines that are parallel to velocity vectors projected onto a
$z=$const plane\cite{BissetHuntRogers2002}}.
As for the inflection points, it is possible that they will
be observed at the interfaces between alternating positive and negative streamwise velocity fluctuations (near-wall streaks),
resulting in a distortion of the velocity profile.

Figures \ref{fig: SNB-800-900-1000}--\ref{fig: 1900_SNB2} give an additional view of the flow topological structure during the self-sustained state of puffs before and after the abrupt onset of exponential decaying to laminar flow.
Besides the streaks visualized by the mapping of streamwise velocity fluctuations, these figures display the sectional streamlines and locations of inflection points.

\begin{figure*}
\centerline{
  \hbox{
    \resizebox{55mm}{!}
    %{\includegraphics[width=0.7 \textwidth]{str2_z-4D_t800}}
    {\includegraphics[width=0.7 \textwidth]{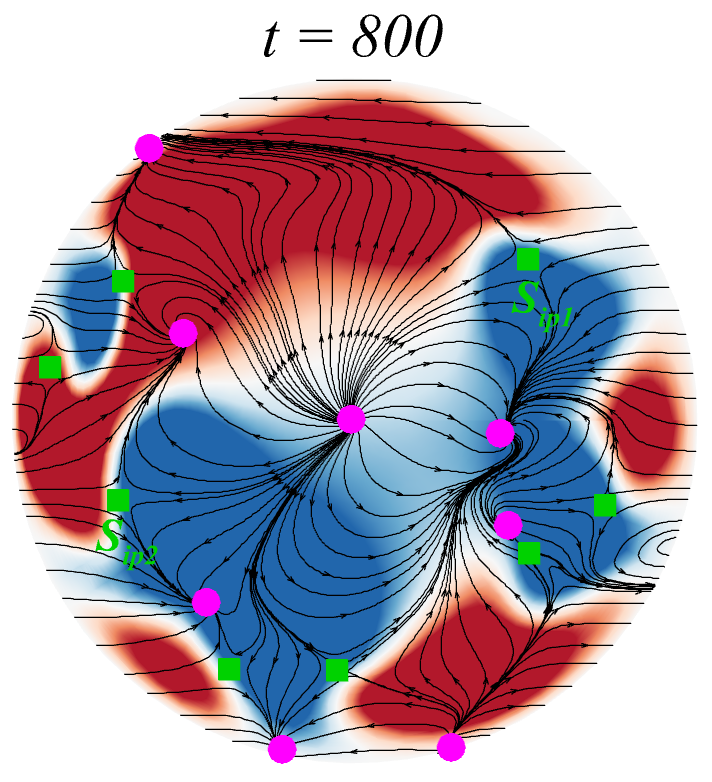}}
    \hspace{1mm}
    \resizebox{55mm}{!}
    %{\includegraphics[width=0.7 \textwidth]{str2_z-4D_t900}}
    {\includegraphics[width=0.7 \textwidth]{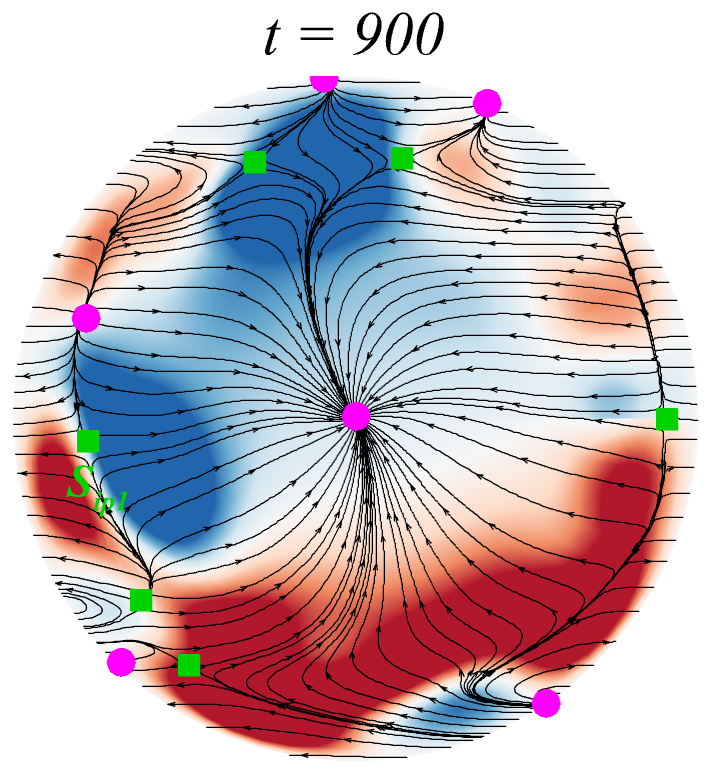}}
    \hspace{1mm}
    \resizebox{55mm}{!}
    %{\includegraphics[width=0.7 \textwidth]{str2_z-4D_t1000}}
    {\includegraphics[width=0.7 \textwidth]{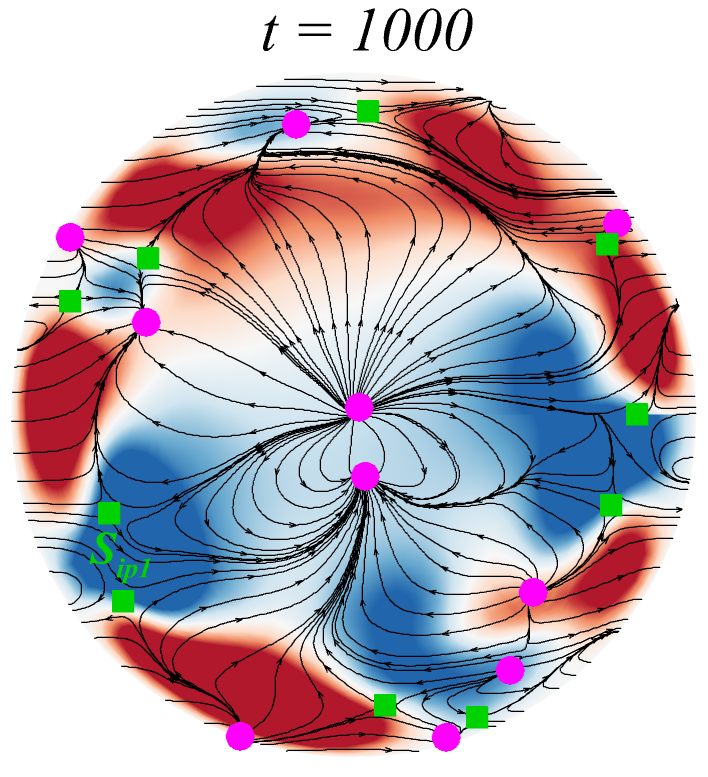}}
  }
}
    \caption{
$Re=1900$, $z=-4D$.  Typical sectional streamlines ($u_r, u_\theta$), along with color mappings of longitudinal velocity
fluctuations ($u_z$); time ($t$) in $D/U_m$ units. Squares (green) and circles (purple) denote saddle and nodal points,
respectively.
}
        \label{fig: SNB-800-900-1000}
\end{figure*}

Figure \ref{fig: SNB-800-900-1000} shows sectional streamlines
of the in-plane projected velocity vectors, ($u_r, u_\theta$), at three time instances within the self-sustaining, statistical
steady-state stage (figure \ref{fig: 1880-1900-1920 ez with tauD}b).
It provides a two-dimensional cut of a three-dimensional motion through a cross-sectional plane. The nodal points at the center (0, 0) are the outcome of projecting the three-dimensional flow onto the cross-sectional plane. In the in-plane velocity field, streamlines converge in one direction and diverge in another, forming a saddle (stagnation) point; nodal points are also stagnation points, but the streamlines move either towards the point or away from it\cite{Perry1987}.
\begin{figure*}
\centerline{
  \hbox{
    \resizebox{50mm}{!}
    %{\includegraphics[width=0.7 \textwidth]{E:/Papers/TurboLam/PRF/Figures/str4_z-4D_t1090}}
    {\includegraphics[width=0.7 \textwidth]{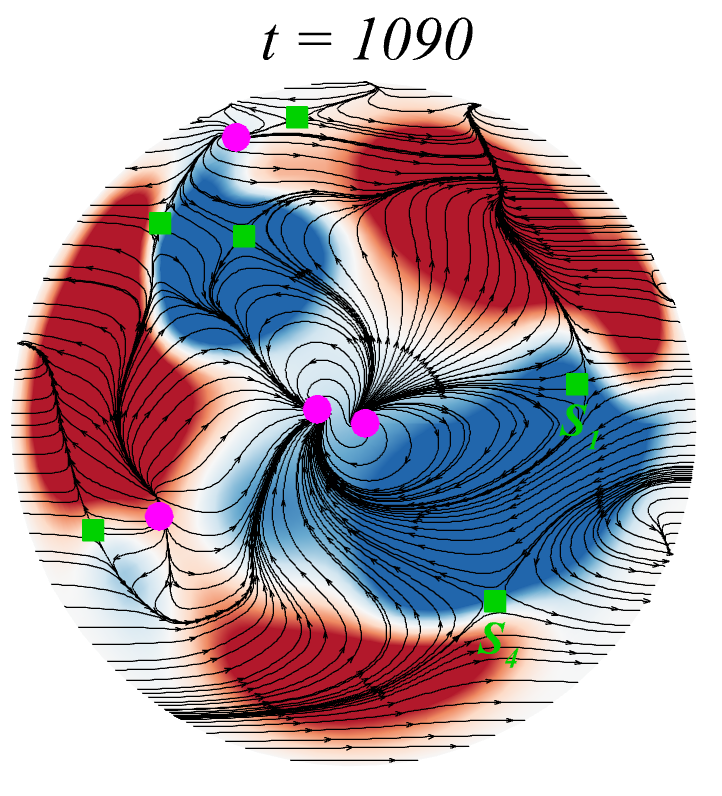}}
    \hspace{1mm}
    \resizebox{50mm}{!}
    %{\includegraphics[width=0.7 \textwidth]{E:/Papers/TurboLam/PRF/Figures/str4_z-4D_t1091}}
    {\includegraphics[width=0.7 \textwidth]{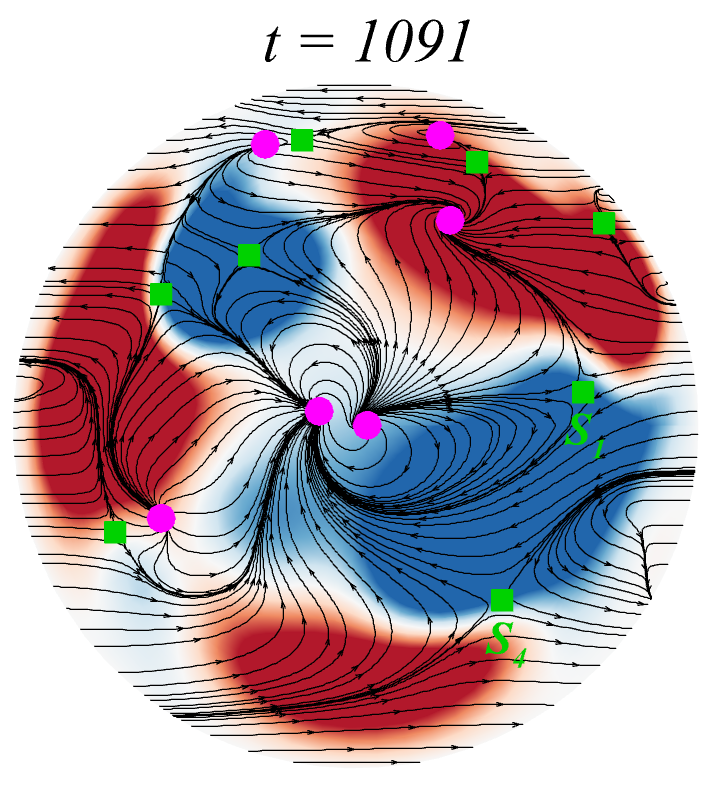}}
    \hspace{1mm}
    \resizebox{50mm}{!}
    %{\includegraphics[width=0.7 \textwidth]{E:/Papers/TurboLam/PRF/Figures/N_str4_z-4D_t1092}}
    {\includegraphics[width=0.7 \textwidth]{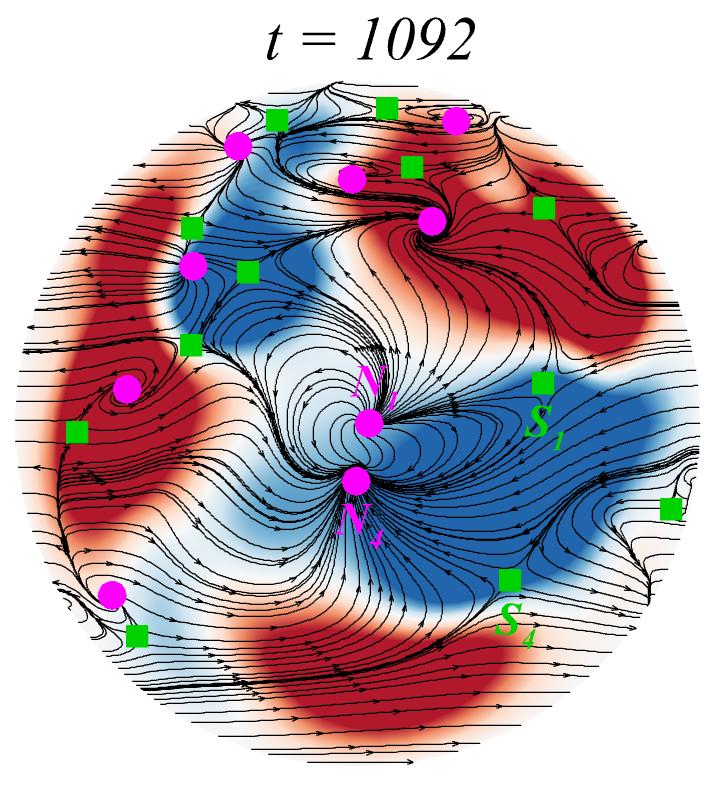}}
  }
}
\centerline{
  \hbox{
    \resizebox{50mm}{!}
    %{\includegraphics[width=0.7 \textwidth]{E:/Papers/TurboLam/PRF/Figures/N_str4_z-4D_t1093}}
    {\includegraphics[width=0.7 \textwidth]{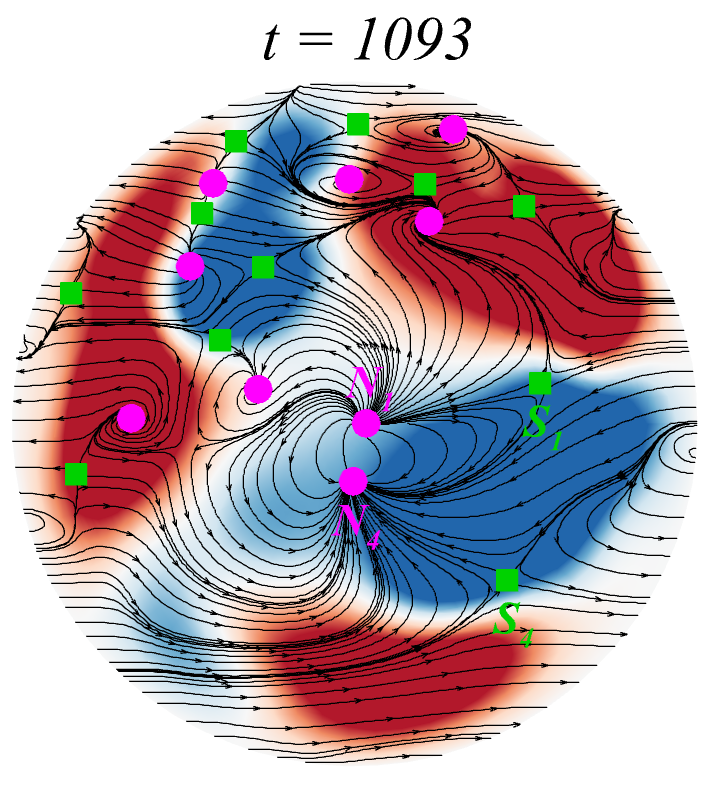}}
    \hspace{1mm}
    \resizebox{50mm}{!}
    %{\includegraphics[width=0.7 \textwidth]{E:/Papers/TurboLam/PRF/Figures/N_str4_z-4D_t1094}}
    {\includegraphics[width=0.7 \textwidth]{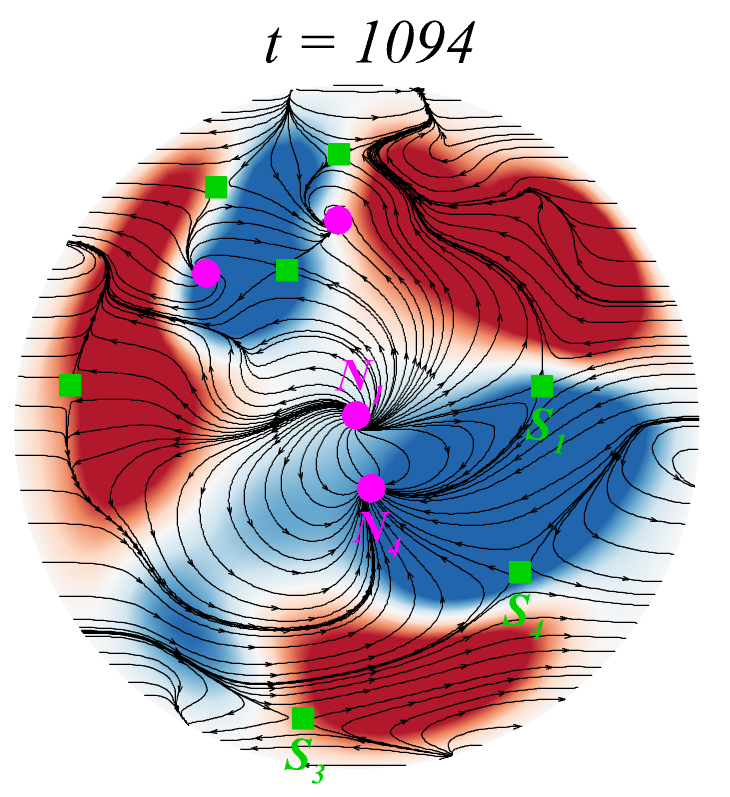}}
    \hspace{1mm}
    \resizebox{50mm}{!}
    %{\includegraphics[width=0.7 \textwidth]{E:/Papers/TurboLam/PRF/Figures/N_str4_z-4D_t1095}}
    {\includegraphics[width=0.7 \textwidth]{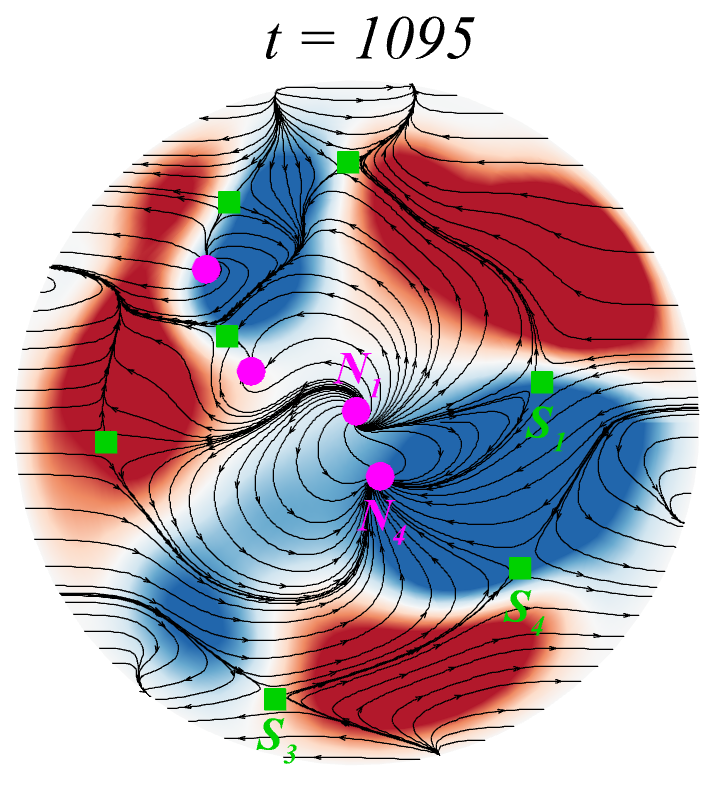}}
  }
}
\centerline{
  \hbox{
    \resizebox{50mm}{!}
    %{\includegraphics[width=0.7 \textwidth]{E:/Papers/TurboLam/PRF/Figures/N_str4_z-4D_t1096}}
    {\includegraphics[width=0.7 \textwidth]{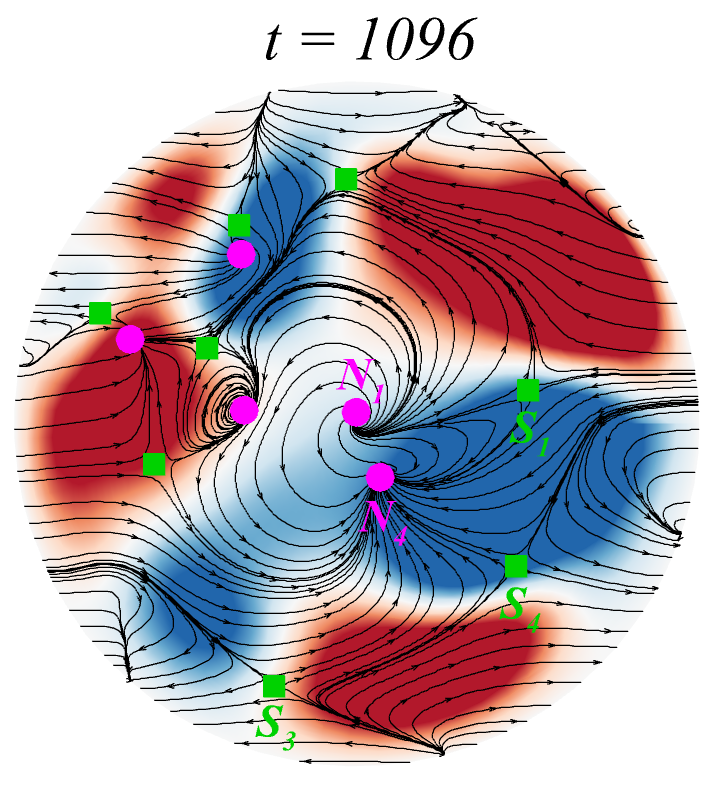}}
    \hspace{1mm}
    \resizebox{50mm}{!}
    %{\includegraphics[width=0.7 \textwidth]{E:/Papers/TurboLam/PRF/Figures/N_str4_z-4D_t1097}}
    {\includegraphics[width=0.7 \textwidth]{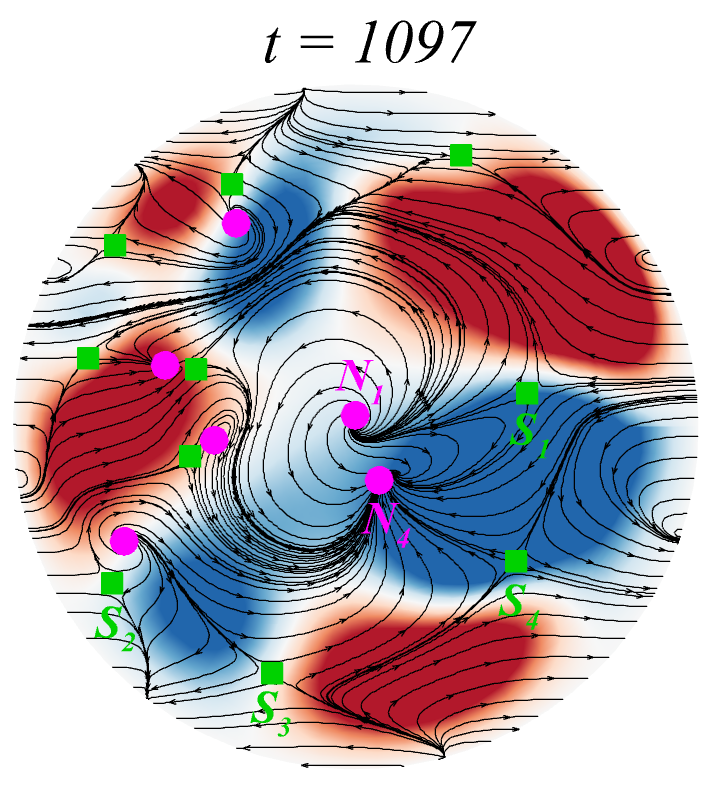}}
    \hspace{1mm}
    \resizebox{50mm}{!}
    %{\includegraphics[width=0.7 \textwidth]{E:/Papers/TurboLam/PRF/Figures/N_str4_z-4D_t1098}}
    {\includegraphics[width=0.7 \textwidth]{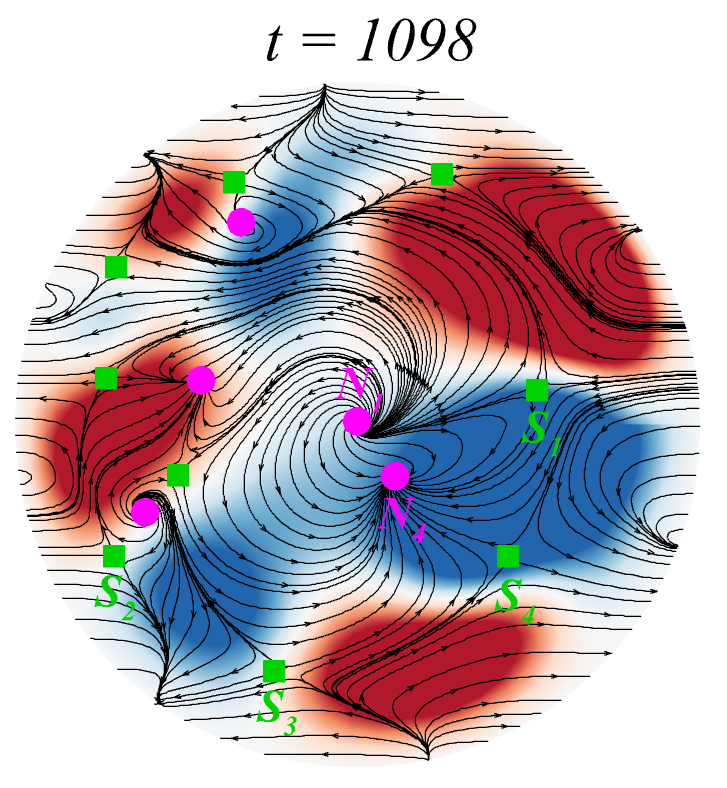}}
  }
}
\centerline{
  \hbox{
    \resizebox{50mm}{!}
    %{\includegraphics[width=0.7 \textwidth]{E:/Papers/TurboLam/PRF/Figures/N_str4_z-4D_t1099}}
    {\includegraphics[width=0.7 \textwidth]{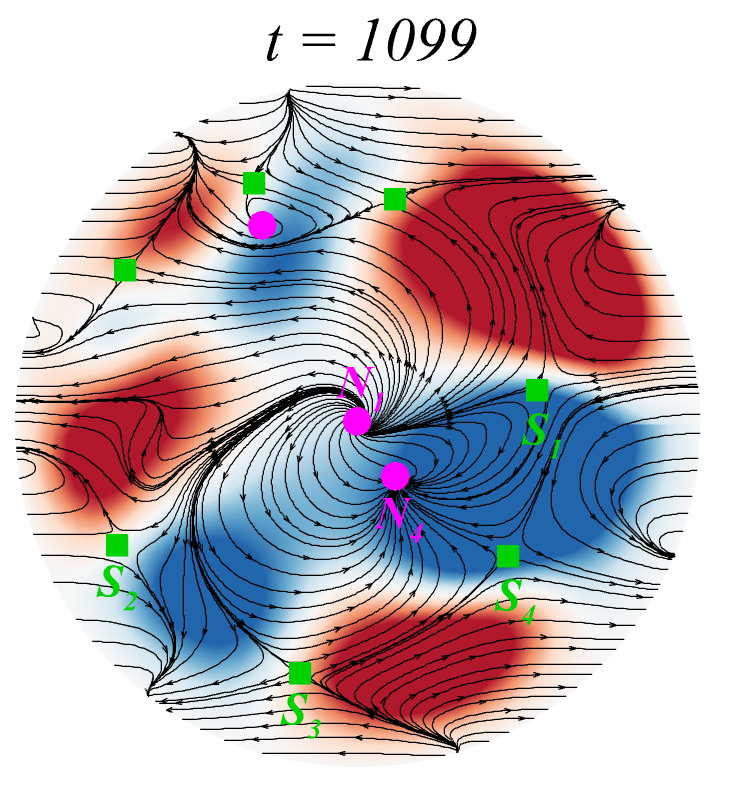}}
    \hspace{1mm}
    \resizebox{50mm}{!}
    %{\includegraphics[width=0.7 \textwidth]{E:/Papers/TurboLam/PRF/Figures/N_str4_z-4D_t1100}}
    {\includegraphics[width=0.7 \textwidth]{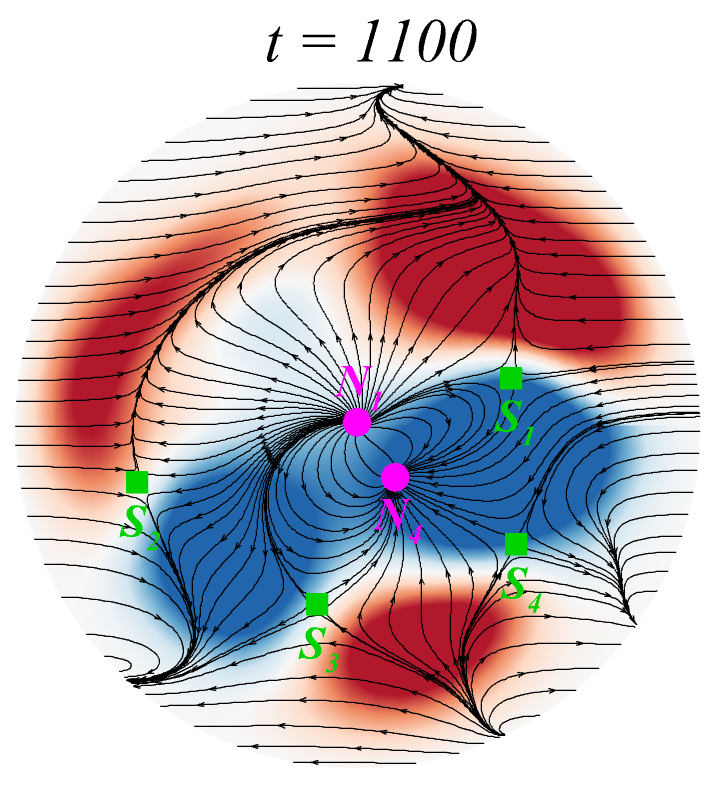}}
    \hspace{1mm}
    \resizebox{50mm}{!}
    %{\includegraphics[width=0.7 \textwidth]{E:/Papers/TurboLam/PRF/Figures/N_str4_z-4D_t1101}}
    {\includegraphics[width=0.7 \textwidth]{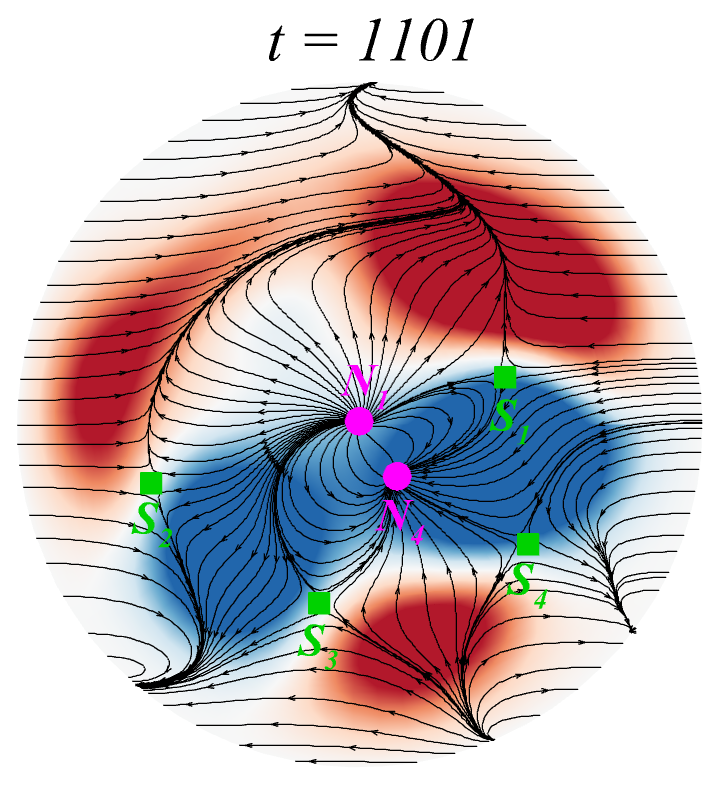}}
  }
}
    \caption{(continued)
}
        \label{fig: 1900_SNB}
\end{figure*}
\begin{figure*}
\centerline{
  \hbox{
    \resizebox{50mm}{!}
    %{\includegraphics[width=0.7 \textwidth]{E:/Papers/TurboLam/PRF/Figures/N_str4_z-4D_t1102}}
    {\includegraphics[width=0.7 \textwidth]{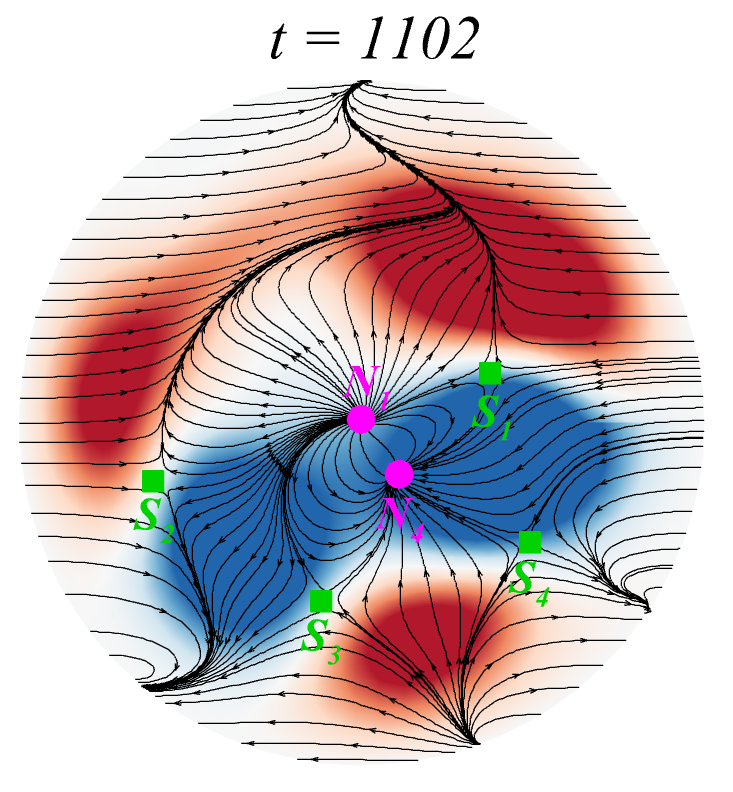}}
    \hspace{1mm}
    \resizebox{50mm}{!}
    %{\includegraphics[width=0.7 \textwidth]{E:/Papers/TurboLam/PRF/Figures/N_str4_z-4D_t1103}}
    {\includegraphics[width=0.7 \textwidth]{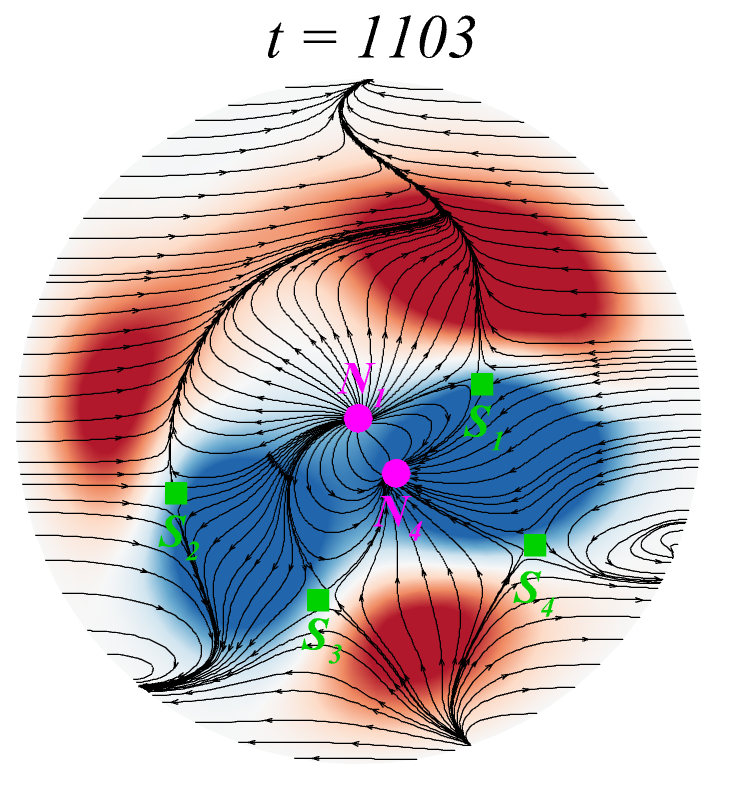}}
    \hspace{1mm}
    \resizebox{50mm}{!}
    %{\includegraphics[width=0.7 \textwidth]{E:/Papers/TurboLam/PRF/Figures/N_str4_z-4D_t1104}}
    {\includegraphics[width=0.7 \textwidth]{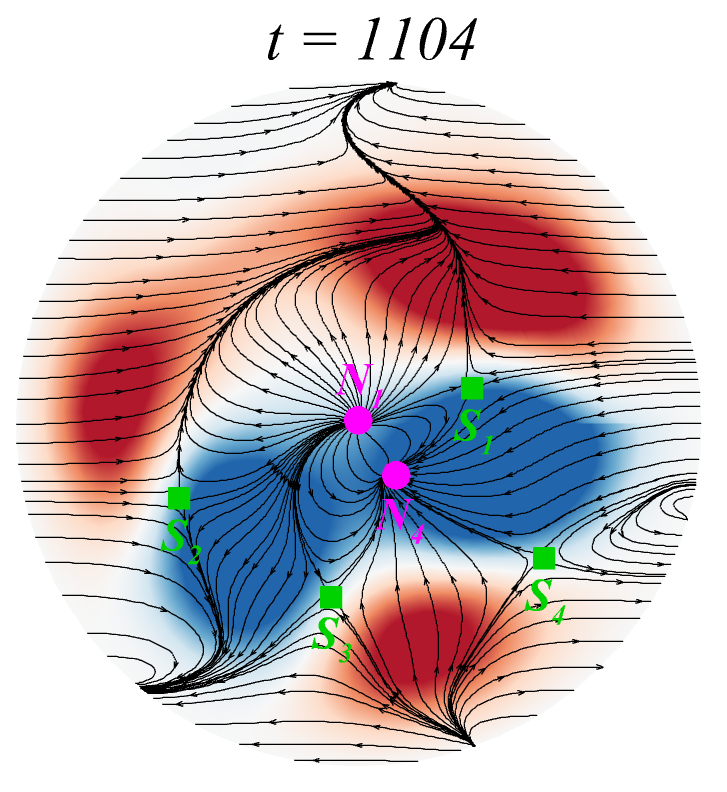}}
  }
}
\centerline{
  \hbox{
    \resizebox{50mm}{!}
    %{\includegraphics[width=0.7 \textwidth]{E:/Papers/TurboLam/PRF/Figures/N_str4_z-4D_t1105}}
    {\includegraphics[width=0.7 \textwidth]{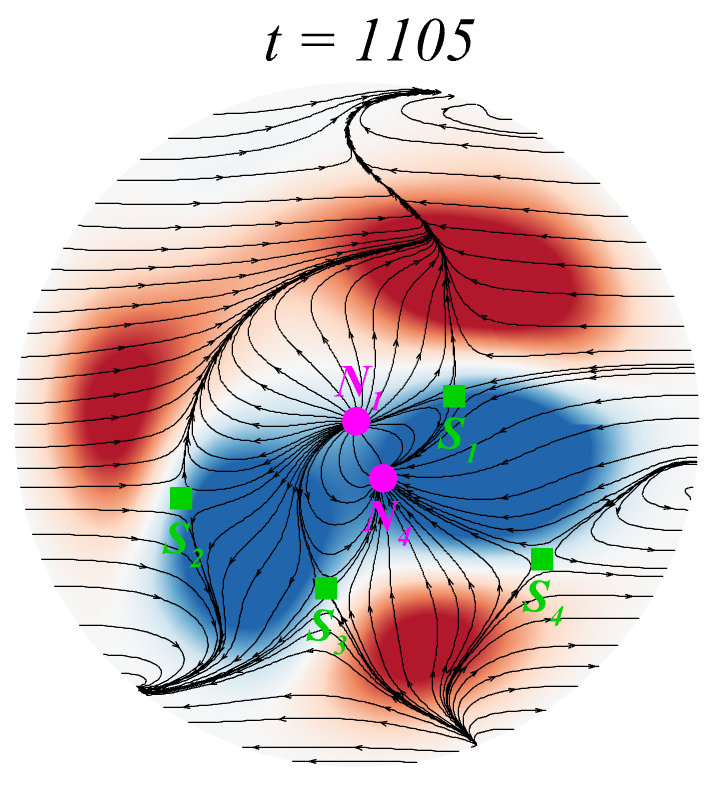}}
    \hspace{1mm}
    \resizebox{50mm}{!}
    %{\includegraphics[width=0.7 \textwidth]{E:/Papers/TurboLam/PRF/Figures/N_str4_z-4D_t1105pt5}}
    {\includegraphics[width=0.7 \textwidth]{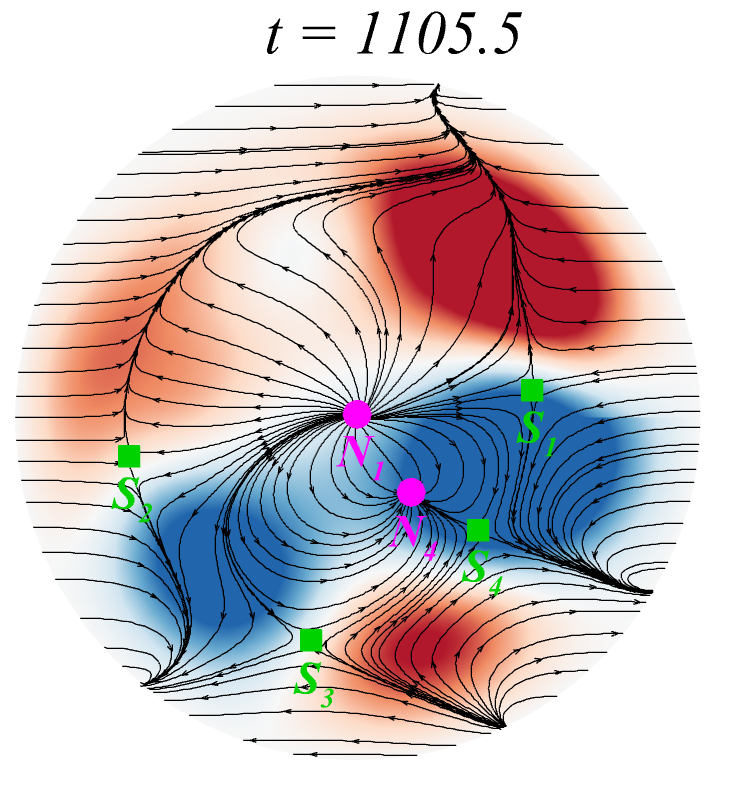}}
    \hspace{1mm}
    \resizebox{50mm}{!}
    %{\includegraphics[width=0.7 \textwidth]{E:/Papers/TurboLam/PRF/Figures/N_str4_z-4D_t1106}}
    {\includegraphics[width=0.7 \textwidth]{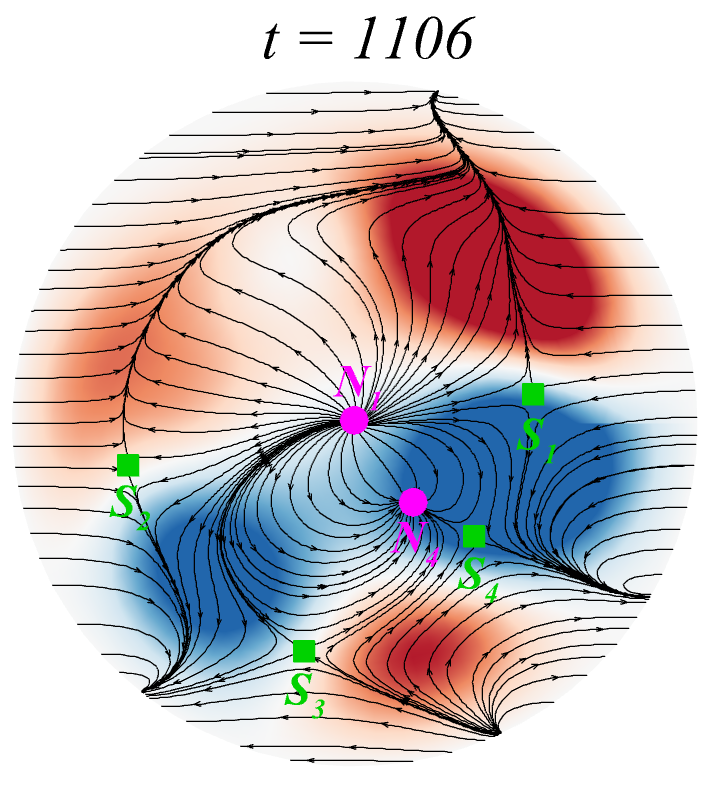}}
  }
}
\centerline{
  \hbox{
    \resizebox{50mm}{!}
    %{\includegraphics[width=0.7 \textwidth]{E:/Papers/TurboLam/PRF/Figures/N_str4_z-4D_t1107}}
    {\includegraphics[width=0.7 \textwidth]{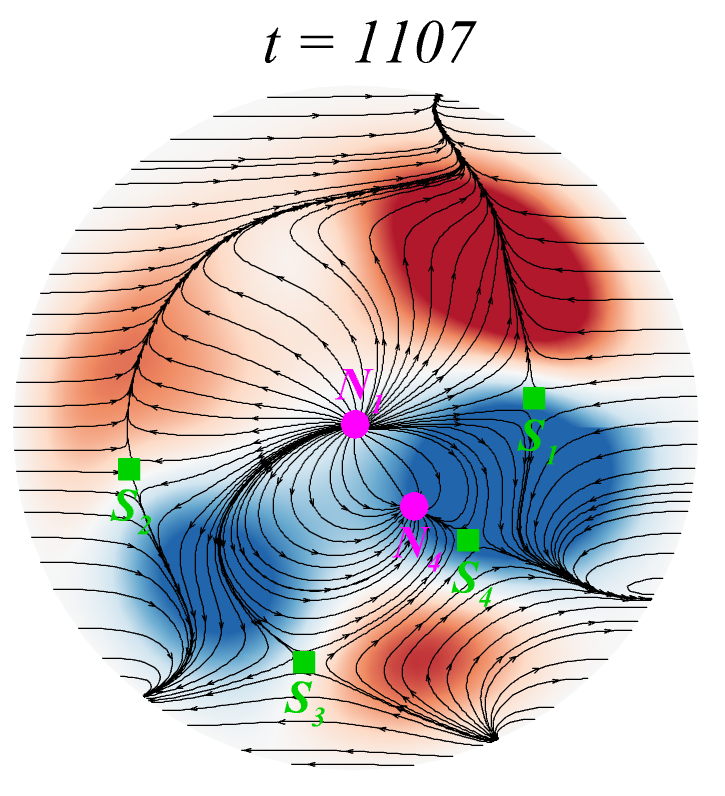}}
    \hspace{1mm}
    \resizebox{50mm}{!}
    %{\includegraphics[width=0.7 \textwidth]{E:/Papers/TurboLam/PRF/Figures/N_str4_z-4D_t1108}}
    {\includegraphics[width=0.7 \textwidth]{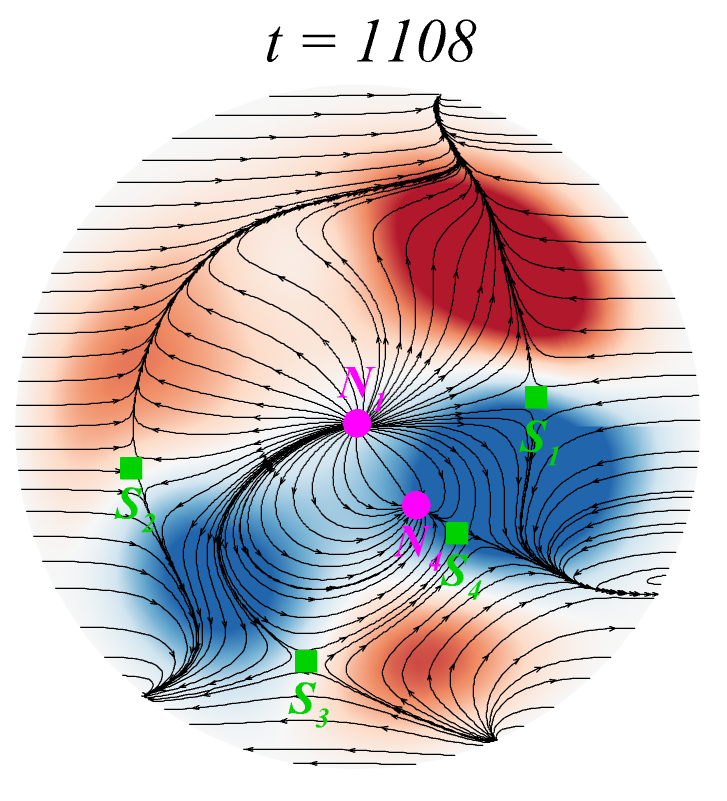}}
    \hspace{1mm}
    \resizebox{50mm}{!}
    %{\includegraphics[width=0.7 \textwidth]{E:/Papers/TurboLam/PRF/Figures/N_str4_z-4D_t1109}}
    {\includegraphics[width=0.7 \textwidth]{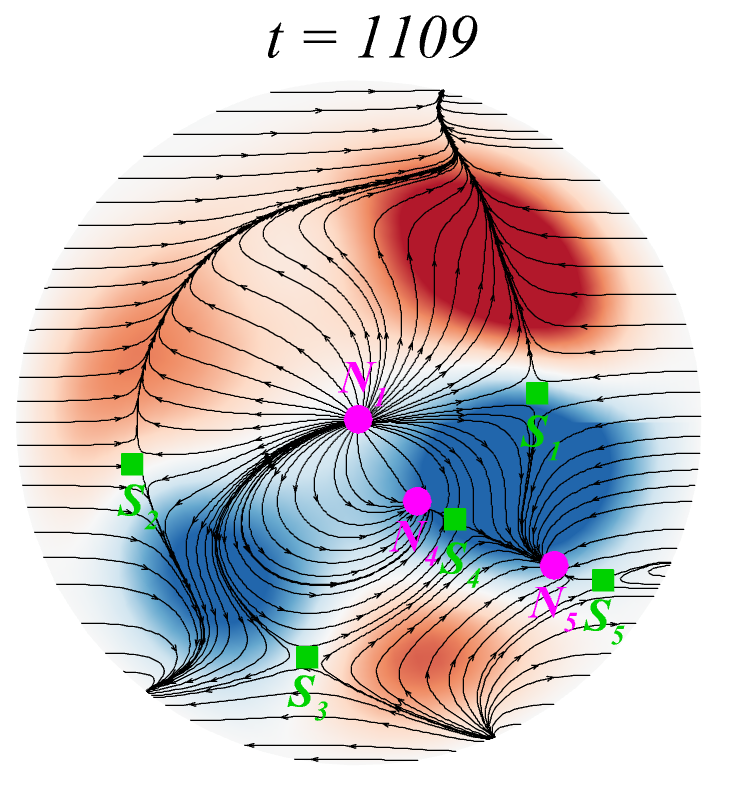}}
  }
}
\centerline{
  \hbox{
    \resizebox{50mm}{!}
    %{\includegraphics[width=0.7 \textwidth]{E:/Papers/TurboLam/PRF/Figures/N_str4_z-4D_t1110}}
    {\includegraphics[width=0.7 \textwidth]{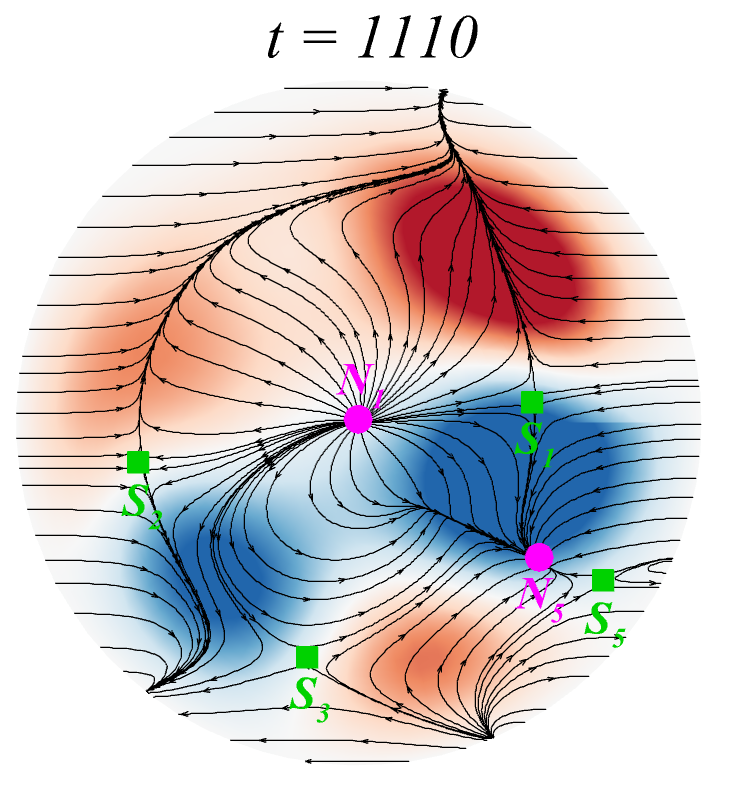}}
    \hspace{1mm}
    \resizebox{50mm}{!}
    %{\includegraphics[width=0.7 \textwidth]{E:/Papers/TurboLam/PRF/Figures/N_str4_z-4D_t1111}}
    {\includegraphics[width=0.7 \textwidth]{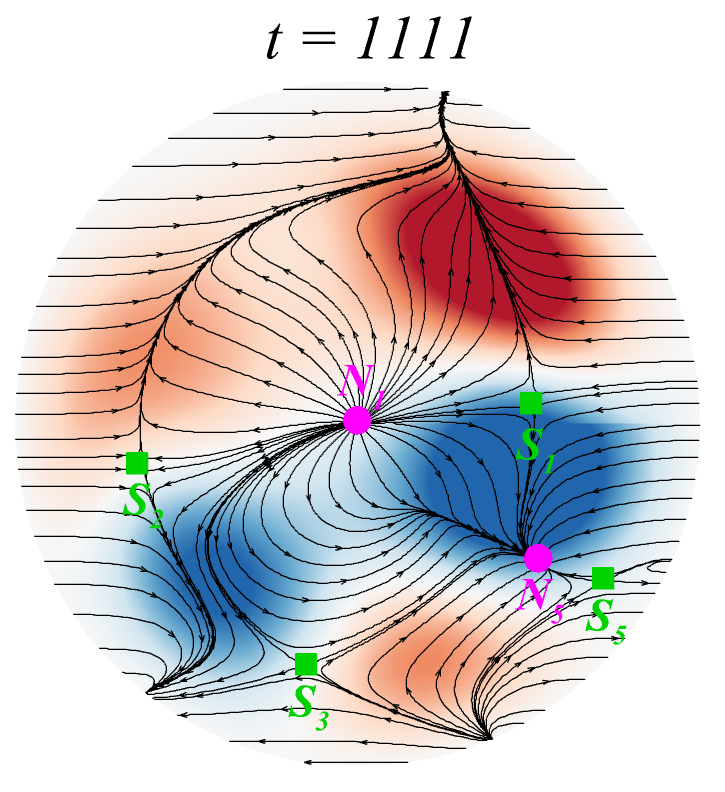}}
    \hspace{1mm}
    \resizebox{50mm}{!}
    %{\includegraphics[width=0.7 \textwidth]{E:/Papers/TurboLam/PRF/Figures/N_str4_z-4D_t1112}}
    {\includegraphics[width=0.7 \textwidth]{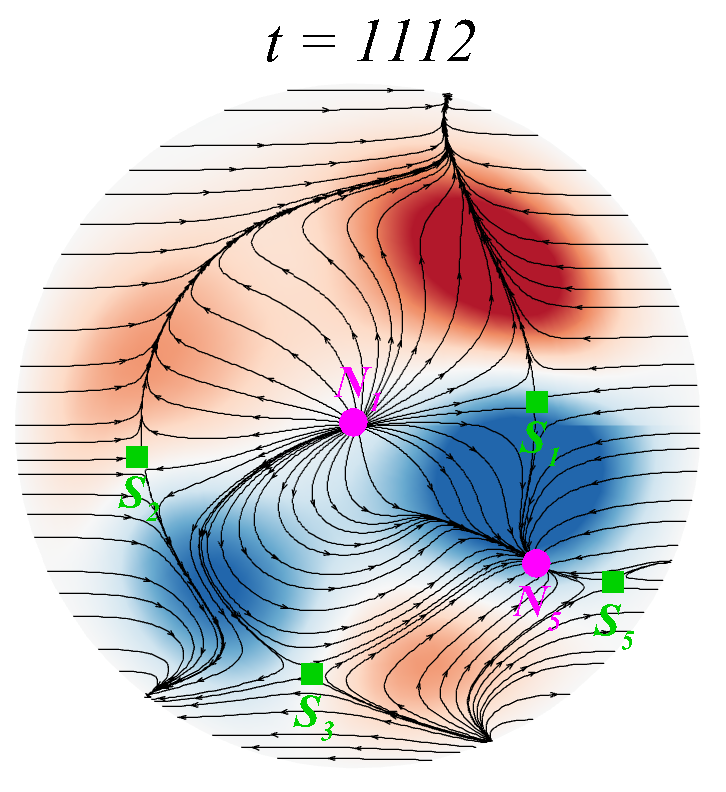}}
  }
}
    \caption{$Re=1900$, initial conditions IC1, $z=-4D$. Sectional phase portraits ($u_r, u_\theta$) and streamwise velocity
fluctuations ($u_z$, color) are shown at different time instants during the decay stage
(figure \ref{fig: 1880-1900-1920 ez with tauD}b) to emphasize local flow structures. Squares (green) and circles (purple)
denote saddle and nodal points, respectively.
}
        \label{fig: 1900_SNB2}
\end{figure*}
In this section, we follow the temporal evolution in the cross-sectional plane $z=-4D$, which is the trailing edge of the turbulent puff, as time counts the changes in the global characteristic, specifically the total turbulent energy ($e_z$) accumulated in the puff. The evolution of the near-wall structure can be traced by following the sectional streamlines.
Sectional streamlines and streamwise velocity fluctuations are shown in figure \ref{fig: 1900_SNB} at various time instances during the decay stage. Before the onset of decay ($t=1100$), denoted by $\bigoplus$ sign in figures \ref{fig: 1880-1900-1920 ez with tauD}b and \ref{fig: ez-vs-D-1880-1900-1920}c, numerous saddle-node pairs, $S/N$, are observed in the near-wall region. Some of them persisted for quite a long time, others existed only for a short period of time. Their distribution is irregular, they can occur at any point in the cross-section and eventually disappear over time. However, we want to draw attention to the time interval from $t=1089$ till $t=1097$, marked with a marker $\bullet$ in the left inset in figure
\ref{fig: 1880-1900-1920 ez with tauD}b. This time interval is interesting for the last local increase in the energy, $e_z$, before the onset of exponential decay at $t=1100$. The number of saddle points on the $t=1090$ panel (figure \ref{fig: 1900_SNB}) is relatively small, although on the $t=1088$ panel (figure \ref{fig: 1900_z-4D-z0-1088-1100-1112}) there are even fewer.
Particular attention should be given to the significant increase in the number of $S/N$ pairs on the panels of the interval $t=1090$ to $t=1097$ (figure \ref{fig: 1900_SNB}), which is indicative of the local increase in streamwise fluctuations energy, $e_z$, as indicated by the $\bullet$ signs in the left inset of figure \ref{fig: 1880-1900-1920 ez with tauD}b.

From figure \ref{fig: 1900_SNB2}, with the onset of relaminarization ($t>1100$), persistent near-wall structures (streaks)
detach from the wall and move toward the central part of the pipe, decreasing in size, and short-term $S/N$ pairs gradually
disappear.  We identify two saddle–node pairs, $S_4/N_4$ and $S_5/N_5$, that progressively approach each other until
they collide and vanish.

\begin{figure*}
\centerline{{\large \bf(z=-4D)}
  \hbox{
    \resizebox{50mm}{!}
    %{\includegraphics[width=0.7 \textwidth]{fig9a}}
    {\includegraphics[width=0.7 \textwidth]{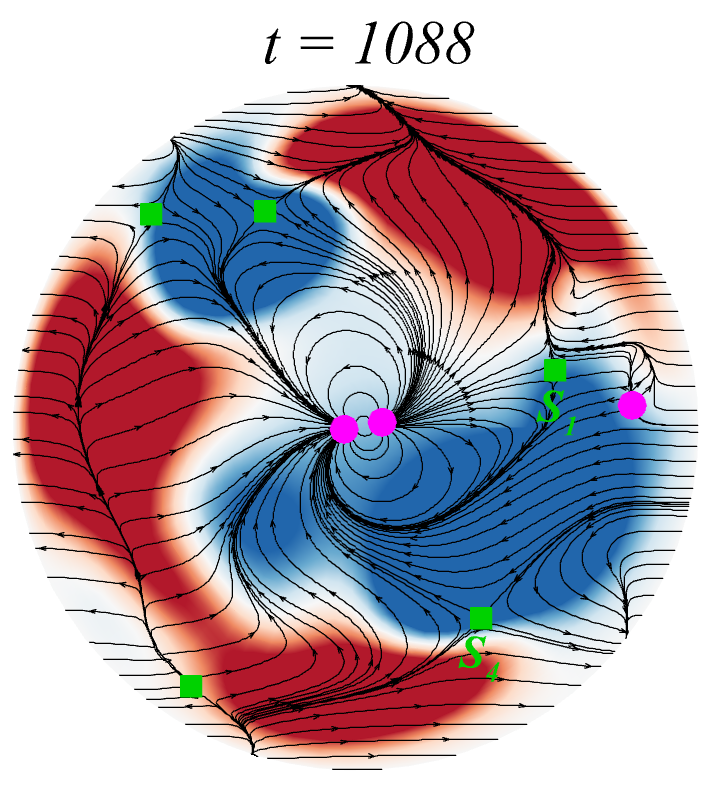}}
    \hspace{1mm}
    \resizebox{50mm}{!}
    %{\includegraphics[width=0.7 \textwidth]{N_str4_z-4D_t1100}}
    {\includegraphics[width=0.7 \textwidth]{fig9_t1100}}
    \hspace{1mm}
    \resizebox{50mm}{!}
     %{\includegraphics[width=0.7 \textwidth]{N_str4_z-4D_t1112}}
     {\includegraphics[width=0.7 \textwidth]{fig10_t1112}}
  }
}
\centerline{{\large \bf(z=0)}\hspace{5mm}
  \hbox{
    \resizebox{50mm}{!}
    %{\includegraphics[width=0.7 \textwidth]{str_z0_t1088}}
    {\includegraphics[width=0.7 \textwidth]{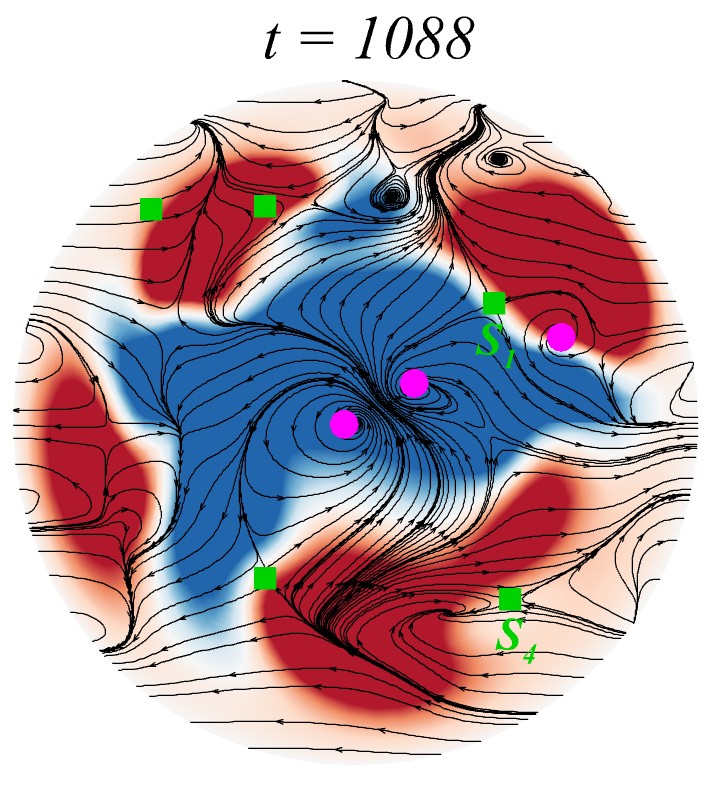}}
    \hspace{1mm}
    \resizebox{50mm}{!}
    %{\includegraphics[width=0.7 \textwidth]{str_z0_t1100}}
    {\includegraphics[width=0.7 \textwidth]{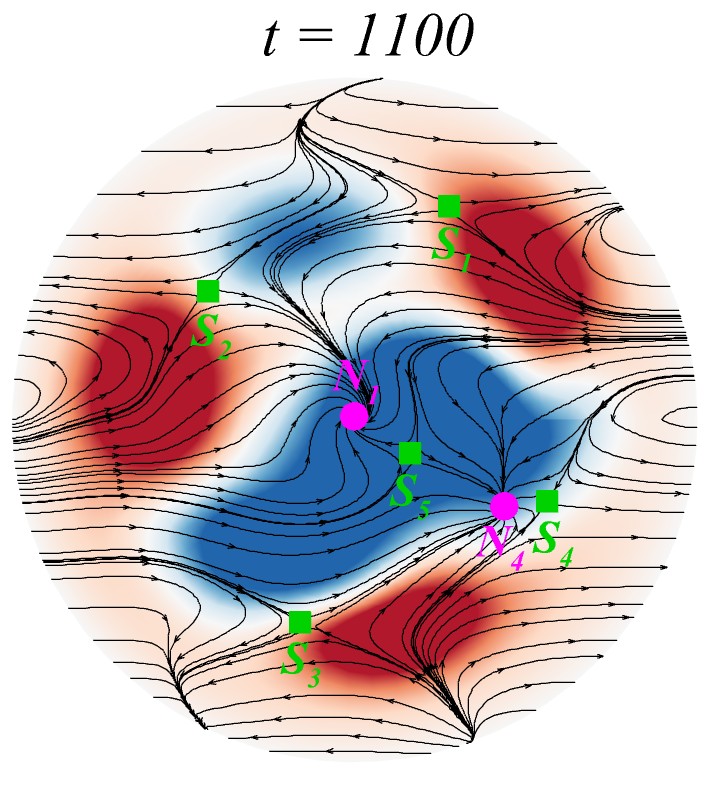}}
    \hspace{1mm}
    \resizebox{50mm}{!}
    %{\includegraphics[width=0.7 \textwidth]{str_z0_t1112}}
    {\includegraphics[width=0.7 \textwidth]{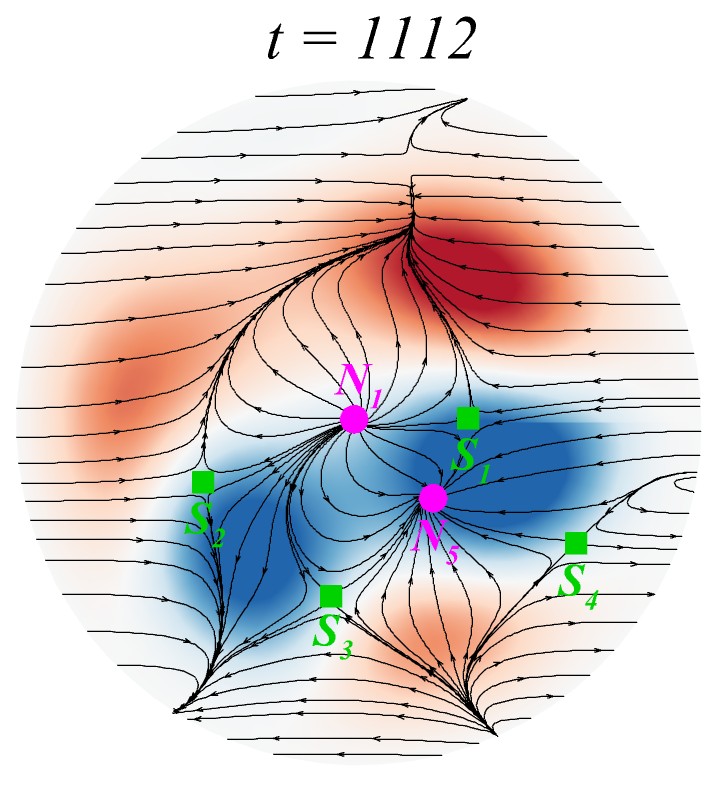}}
  }
}
    \caption{
$Re=1900$, initial conditions IC1. Sectional streamlines ($u_r, u_\theta$) at $z = -4D$ - the laminar-turbulent interface, and at $z=0$ - the location of maximal transverse (turbulent) energy.
}
        \label{fig: 1900_z-4D-z0-1088-1100-1112}
\end{figure*}
Figure \ref{fig: 1900_z-4D-z0-1088-1100-1112} shows the streamlines at $z = -4D$ -- the laminar-turbulent interface, and at
$z=0$ -- the location of maximal transverse (turbulent) energy. Three representative times are selected: before the onset of decay
($t = 1088$), the onset of decay ($t = 1100$), and the halfway point ($t = 1112$) to the complete relaminarization ($t = 1220$),
(figure \ref{fig: 1880-1900-1920 ez with tauD}b). The similarity of the images taken at different spatial locations
but at the same time after entering the relaminarization stage ($t = 1100$) is obvious. This is because the decay is
viscous, which implies that it is governed by a parabolic differential equation.
\begin{figure*}
\centerline{
  \hbox{
    \resizebox{55mm}{!}
    %{\includegraphics[width=0.7 \textwidth]{Uz-at_z-4d_Re1900_t1088d}}
    %{\includegraphics[width=0.7 \textwidth]{fig9d}}
    {\includegraphics[width=0.7 \textwidth]{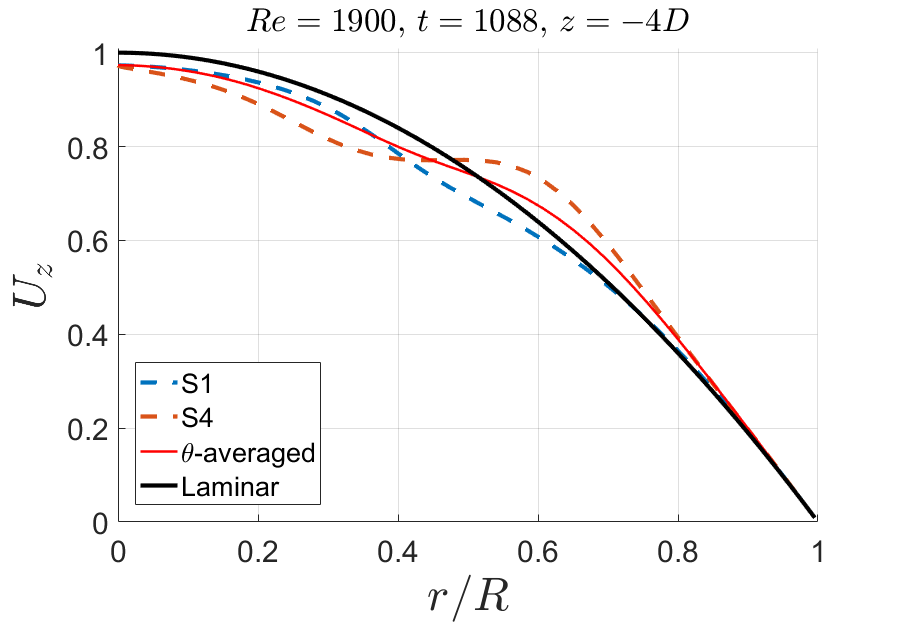}}
    \hspace{1mm}
    \resizebox{55mm}{!}
    %{\includegraphics[width=0.7 \textwidth]{Uz-at_z-4d_Re1900_t1100d}}
    %{\includegraphics[width=0.7 \textwidth]{fig9e}}
    {\includegraphics[width=0.7 \textwidth]{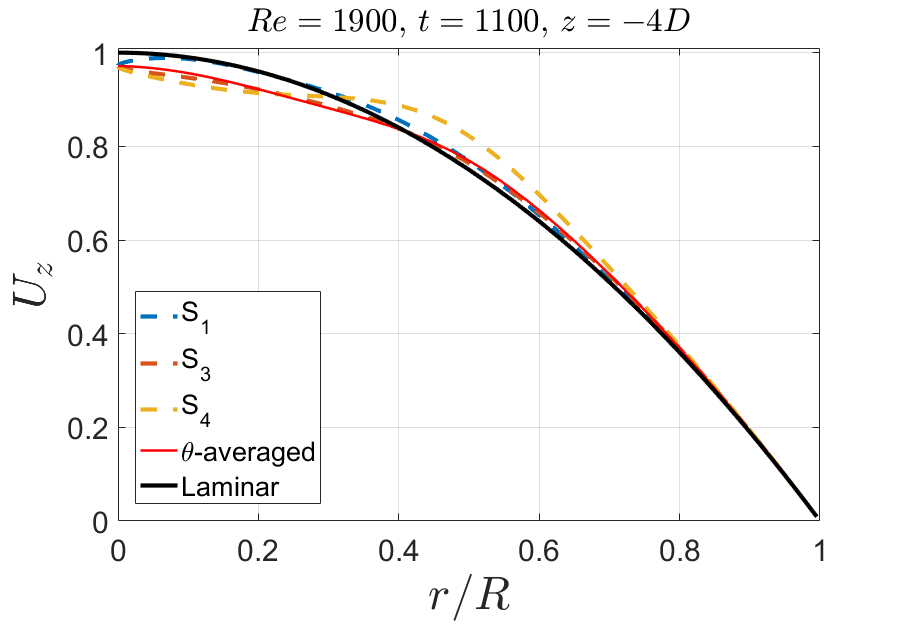}}
    \hspace{1mm}
    \resizebox{55mm}{!}
    %{\includegraphics[width=0.7 \textwidth]{Uz-at_z-4d_Re1900_t1112d}}
    %{\includegraphics[width=0.7 \textwidth]{fig9f}}
    {\includegraphics[width=0.7 \textwidth]{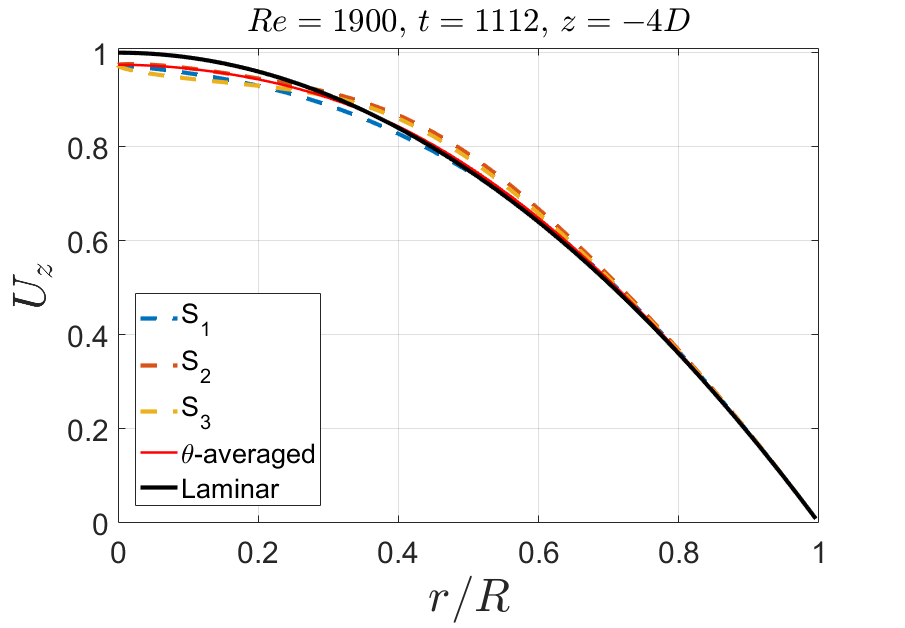}}
  }
}
%%%%%%%%%%%%%%%%%%%%
    \caption{
$Re=1900$, $z=-4D$, IC1. Streamwise velocity ($U_z$) profiles at different circumferential locations; $t$ in $D/U_m$ units.
}
        \label{fig: z-4D-Uz-1088-1100-1112}
\end{figure*}
Figure \ref{fig: z-4D-Uz-1088-1100-1112} displays the corresponding longitudinal velocity
profiles ($U_z$) at different circumferential locations. From figure \ref{fig: z-4D-Uz-1088-1100-1112}, at $t = 1100$, the strength of the inflection points becomes barely visible, and the centerline velocity decreased slightly to $U_c = 0.97$, suggesting a nearly laminar velocity profile at $z = -4D$ and at $z = 0$. Apparently, it is relaminarized as a single unit.
By this point, the total energy is $e_z\approx 0.2$ (figure \ref{fig: 1880-1900-1920 ez with tauD}b), which is quite substantial.
The puff will travel downstream for approximately 100 pipe diameters until complete relaminarization occurs at $t=1220$. The latter is consistent with the experiments reported Ref.~\onlinecite{Kuhnen2018a}.
\onlinecite{Hof2010},
\begin{figure*}
\centerline{
  \hbox{
    \resizebox{50mm}{!}
    %{\includegraphics[width=0.7 \textwidth]{N_str4_z-4D_t7253}}
    {\includegraphics[width=0.7 \textwidth]{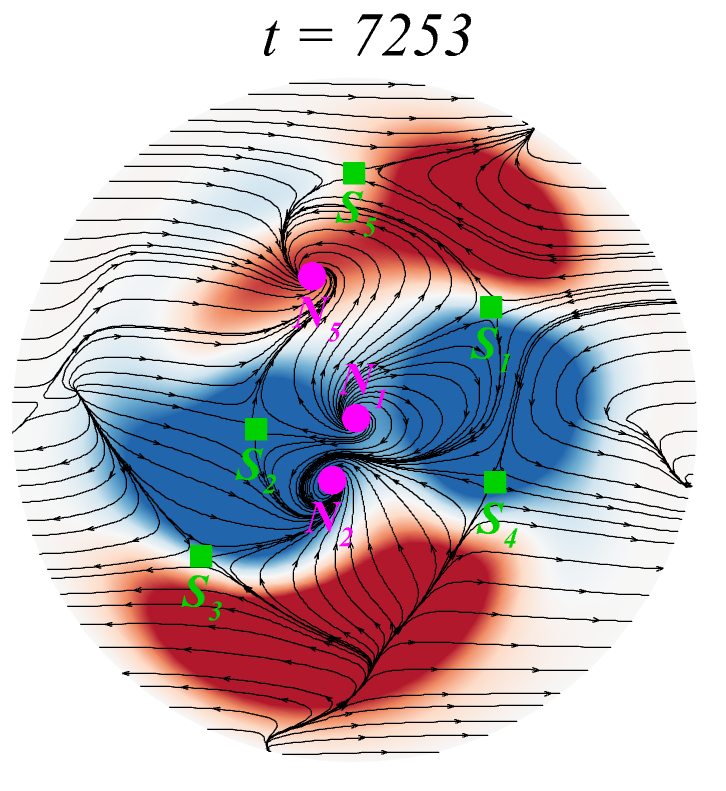}}
    \hspace{1mm}
    \resizebox{50mm}{!}
    %{\includegraphics[width=0.7 \textwidth]{N_str4_z-4D_t7254}}
    {\includegraphics[width=0.7 \textwidth]{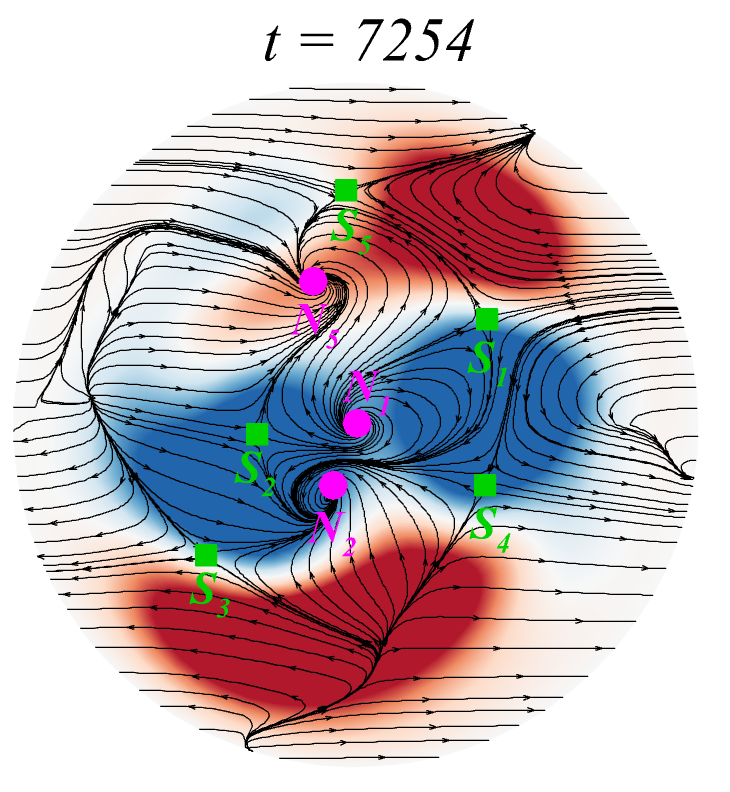}}
    \hspace{1mm}
    \resizebox{50mm}{!}
    %{\includegraphics[width=0.7 \textwidth]{N_str4_z-4D_t7255}}
    {\includegraphics[width=0.7 \textwidth]{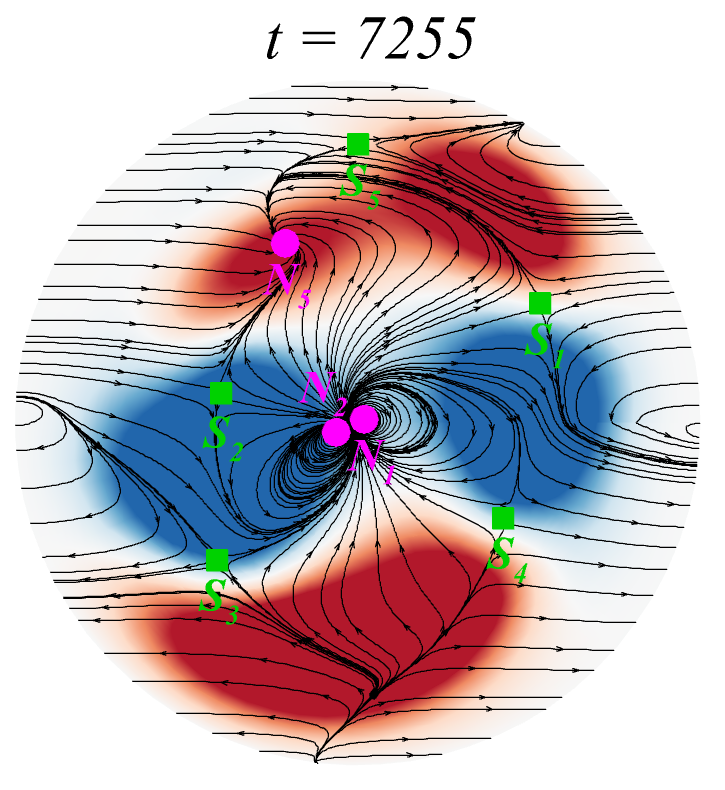}}
  }
}
\centerline{
  \hbox{
    \resizebox{50mm}{!}
    %{\includegraphics[width=0.7 \textwidth]{N_str4_z-4D_t7257}}
    {\includegraphics[width=0.7 \textwidth]{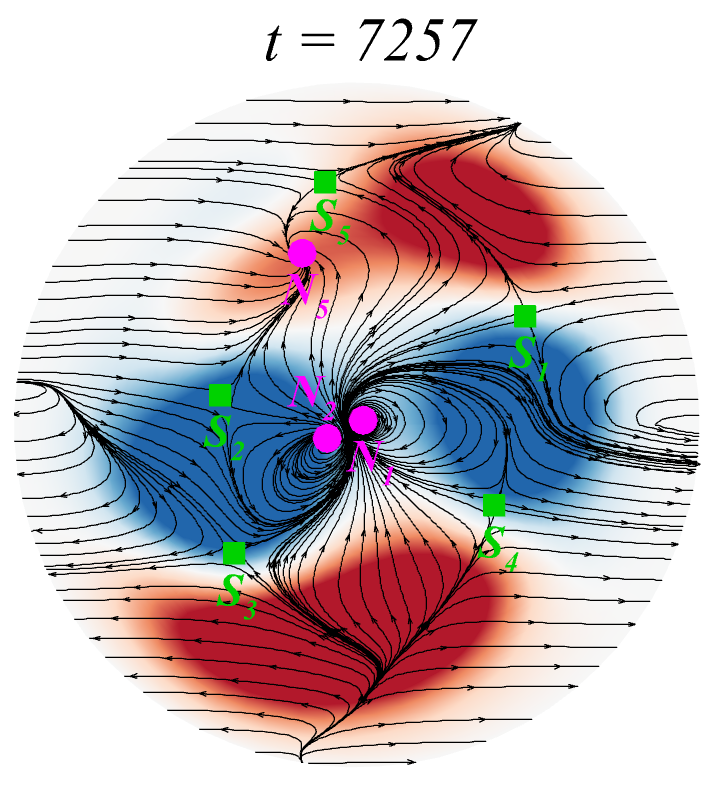}}
    \hspace{1mm}
    \resizebox{50mm}{!}
    %{\includegraphics[width=0.7 \textwidth]{N_str4_z-4D_t7258pt5}}
    {\includegraphics[width=0.7 \textwidth]{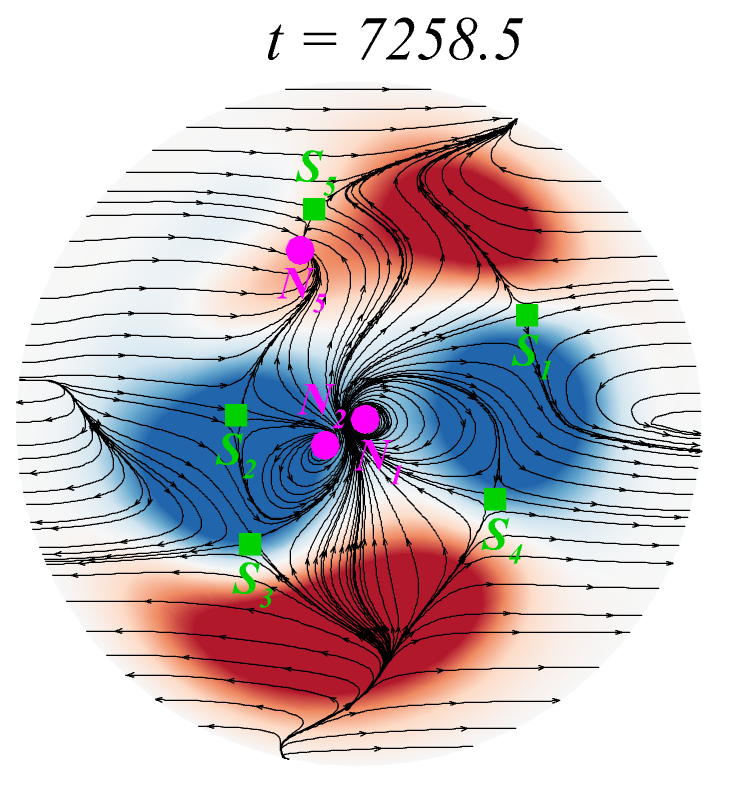}}
    \hspace{1mm}
    \resizebox{50mm}{!}
    %{\includegraphics[width=0.7 \textwidth]{N_str4_z-4D_t7260}}
    {\includegraphics[width=0.7 \textwidth]{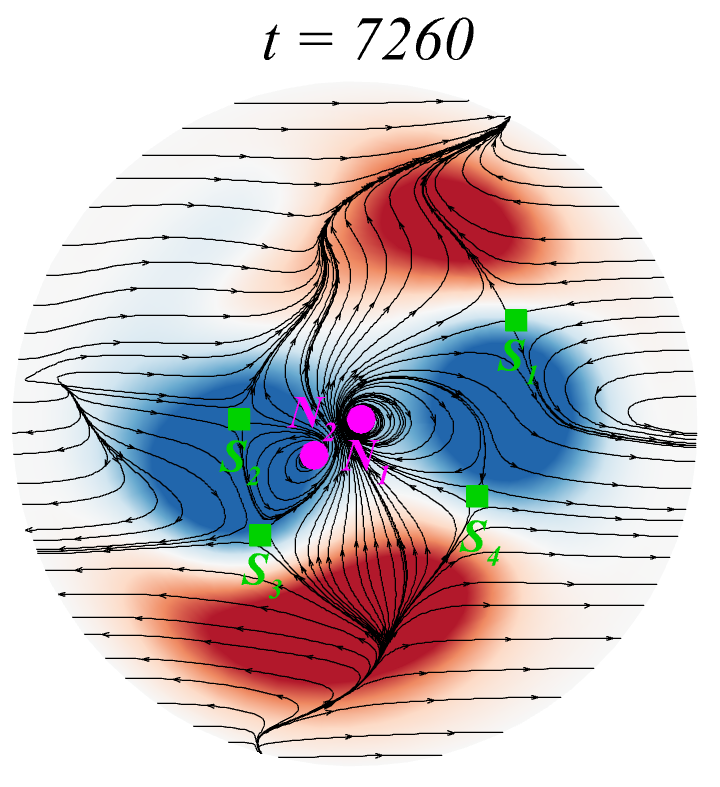}}
  }
}
    \caption{
$Re=1900$, initial conditions IC2, $z=-4D$. Sectional streamlines ($u_r, u_\theta$) and streamwise velocity
fluctuations ($u_z$, color) are shown at different time instants during the decay stage to emphasize local
flow structures. Squares (green) and circles (purple) denote saddle and nodal points, respectively.
}
        \label{fig: 1900_SNBIC2}
\end{figure*}
Figure \ref{fig: 1900_SNBIC2} displays similar patterns to figure \ref{fig: 1900_SNB2}, but it was obtained using different
initial conditions (IC2). The saddle and nodal points of the pair $S_5/N_5$ pair have been observed to move toward each other until they collide and disappear.

\begin{figure*}
{\centering (a) \hspace{7cm}(b)}
\centerline{
  \hbox{
    \resizebox{65mm}{!}
    %{\includegraphics[width=0.7 \textwidth]{1900-IC1_SNB_S4N4d}}
    {\includegraphics[width=0.7 \textwidth]{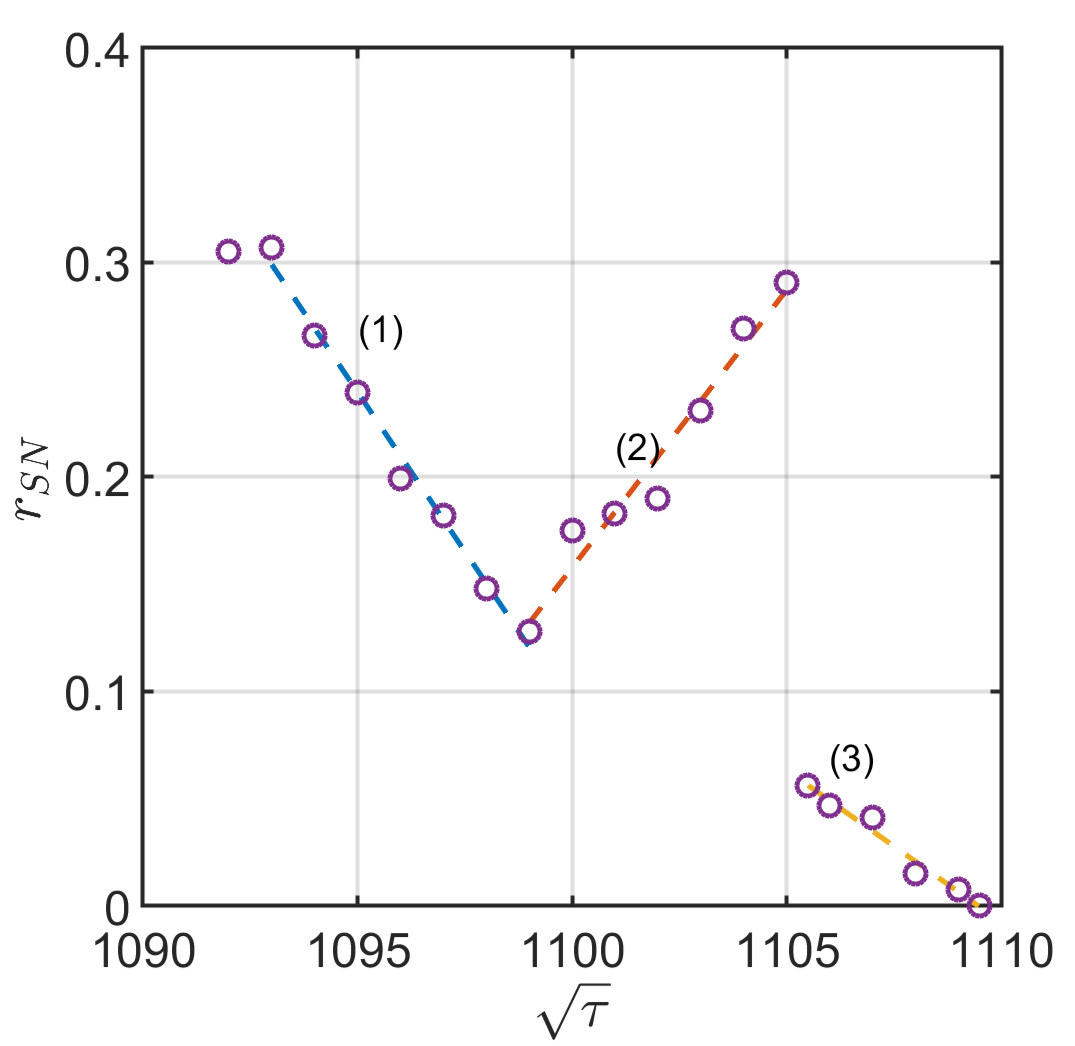}}
    \hspace{10mm}
    \resizebox{65mm}{!}
    %{\includegraphics[width=0.7 \textwidth]{1900-IC2_SNB_S5N5d}}
    {\includegraphics[width=0.7 \textwidth]{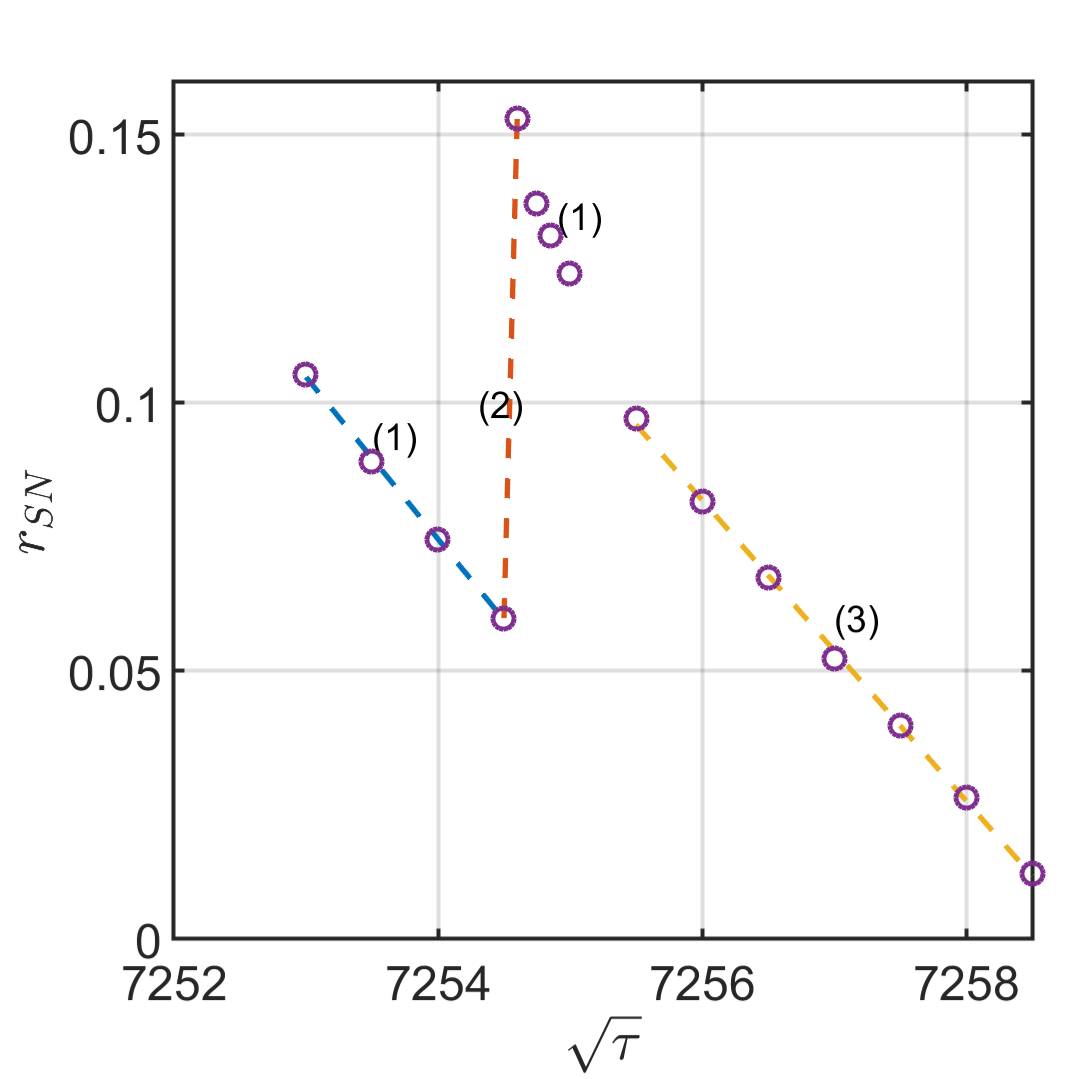}}
  }
}
    \caption{$Re=1900$, $z=-4D$. Three distinct stages for the collision of saddles and nodes: approach (1), repulsion (2) and annihilation at the final stage (3). The square-root scaling $r_{SN} \propto (\tau_*-\tau)^{1/2}$ is valid at the approach and  annihilation stages; (a) $S_4/N_4$ pair in figure \ref{fig: 1900_SNB2},
    (b) $S_5/N_5$ pair in figure \ref{fig: 1900_SNBIC2}.
    }
        \label{fig: SNB-sqrt-law}
\end{figure*}

The described behavior resembles a saddle-node bifurcation developing over time, culminating in annihilation. The topological collapse is not sudden; it develops gradually, suggesting a mechanism like critical slowing down near saddle-node bifurcations of dynamical systems. In figure
\ref{fig: SNB-sqrt-law}, we show the distance between a saddle point and a nodal point, identified as a pair, as a function of the square root of the DNS time of the corresponding snapshot. Straight lines indicate the {\it square-root scaling law}, a generic characteristic of the saddle-node bifurcation\cite{Strogatz2018}: $r_{SN} \propto \sqrt{\tau^{*} - \tau}$.
Here, $r_{SN}$ denotes the distance between a saddle and a node.
%ZZZZZZZZZZZZZZZZZZZZZZZZZZZZZZZZZZZZZZZZZZZZZZZZZZZZZZZ
%\comment{
%} 
\section{CONCLUSIONS}\label{sec: conclusions}
This study explored the relaminarization of turbulent puffs in pipe flow at Reynolds numbers $Re>$1870. In particular, the
exponential energy decay that has been previously observed at $Re$=1720--1870 in Ref. \onlinecite{KhanArogetiYakhot2024} has been confirmed.
Thus, direct numerical simulations carried out across $Re$ = 1720--1920 demonstrated that the energy of streamwise velocity
fluctuations decreases exponentially over time after a sudden break from an equilibrium turbulent state. We have shown that the
characteristic time constant of the decay rate is $\tau_D \approx$ 33--35 in $D/U_m$ units. It has been determined that
the decay rate approximation from Ref. \onlinecite{NarasimhaSreeni1979} is successful when a constant term is incorporated,
$\kappa=\tau_D^{-1}=B(d-Re)^3+C$, where $B=O(10^{-10})$ and $C\approx 0.3$ emphasize the weak $Re$-dependence.

The study revealed
peculiar features of the relaminarization process. The intermittent laminar–turbulent interface was observed at the upstream (trailing) edge of a puff, with sectional streamline patterns that displayed saddle points.
At the trailing edge, the intensity of velocity fluctuations and spatial density of saddle and nodal points decrease over time,
indicating the relaminarization. Inflection points in instantaneous streamwise
velocity radial profiles were identified as indicators of instability, particularly near the trailing edge of a puff.
The weakening of inflection points manifests the collapse of the self-sustaining mechanism of turbulence.

A complex topology was revealed by sectional streamlines, particularly saddle-node pairs that were associated with regions with inflection points in the velocity radial profile.
The saddle and nodal points have been observed in pairs, where they move toward each other until they collide and vanish. An evolving saddle-node bifurcation that ultimately leads to annihilation is reminiscent of this behavior.
During relaminarization, saddle-node pairs that were initially situated near the wall gradually relocated toward the flow core.
Tracing a vanishing pair over time revealed that the distance between a saddle point and a nodal point is a function of the square root of time. This implies the square-root scaling law, a generic characteristic of the saddle-node bifurcation, which is perhaps the simplest and best-known bifurcation in dynamical systems.\\
{\bf Acknowledgement.} This study was supported by the Israel Science Foundation Grant 2228/22.

%%%%%%%%%%%%%%%%%%
%\bibliography{aipsamp}% Produces the bibliography via BibTeX.
\bibliography{refspuffPoF}% Produces the bibliography via BibTeX.

\end{document}